\documentclass[smallextended]{svjour3} 
\usepackage[english]{babel}
\usepackage[utf8]{inputenc}
\usepackage{pgf,tikz,pgfplots}
\usetikzlibrary{arrows}
\usepackage{amssymb}
\usepackage{bm}
\usepackage{csquotes}
\expandafter\let\csname equation*\endcsname\relax
\expandafter\let\csname endequation*\endcsname\relax
\usepackage{amsmath}
\usepackage{amsfonts}
\usepackage{amstext}
\usepackage{amssymb}
\usepackage{physics}
\usepackage{amsmath}
\usepackage{amsfonts}
\usepackage{amssymb}

\usepackage{physics}
\usepackage{color}
\usepackage{hyperref}
\usepackage{enumerate}
\usepackage{graphicx}
\usepackage{float}
\usepackage{amsmath}
\usepackage{amsfonts}
\usepackage{setspace} 
\usepackage{lipsum}
\usepackage{dcolumn}
\usepackage{hyperref}
\usepackage{subfigure}
\usepackage{epsfig}
\usepackage{epsf}
\usepackage{epstopdf}

\usepackage{color}
\usepackage{enumitem}
\usepackage{bm}
\usepackage{subfigure}
\usepackage{xcolor}
\usepackage{tikz}
\usepackage[margin=1in]{geometry}
\usepackage{pgfplots}
\usepackage{esint}

\newcommand\p{\partial}
\newcommand\f{\frac}
\newcommand{\la}{\langle} 
\newcommand{\ra}{\rangle} 
\newcommand{\be}{\begin{equation}}
\newcommand{\ee}{\end{equation}}
\newcommand{\bea}{\begin{eqnarray}}
\newcommand{\eea}{\end{eqnarray}}

\begin{document}

\title{Boltzmann's entropy during free expansion of an interacting ideal gas}

\author{Subhadip Chakraborti\textsuperscript{1,2}, Abhishek Dhar\textsuperscript{1} and Anupam Kundu\textsuperscript{1}}

\institute{${}^{1}$ International Centre for Theoretical Sciences, Bengaluru, India-560089 \\
	${}^{2}$ Friedrich-Alexander-Universit\"{a}t Erlangen-N\"{u}rnberg, Cauerstrasse 11, 91058 Erlangen, Germany \\
	\email{subhadip.chakraborti@fau.de}\\
	\email{abhishek.dhar@icts.res.in}\\
	\email{anupam.kundu@icts.res.in}}

\date{Received: date / Accepted: date}

\maketitle

\begin{abstract}
In this work we study the evolution of Boltzmann's entropy in the context of free expansion of a one dimensional interacting gas inside a box. Boltzmann's entropy is defined for single microstates and is given by the phase-space volume occupied by microstates with the same value of macrovariables which are coarse-grained physical observables. We demonstrate the idea of typicality in the growth of  the Boltzmann's entropy  for two choices of  macro-variables -- the single particle phase space distribution and the hydrodynamic fields. Due to the presence of interaction, the growth curves for both these entropies are observed to converge to a  monotonically increasing limiting curve, on taking the  appropriate order of limits, of large system size and small coarse graining scale. Moreover, we observe that the limiting growth curves for the two choices of entropies are identical as implied by local thermal equilibrium. We also discuss issues related to finite size and finite coarse gaining scale which lead interesting features such as oscillations in the entropy growth curve. We also discuss  shocks observed in the  hydrodynamic fields. 
\end{abstract}

\tableofcontents

\section{Introduction}
\label{sec:intro}
Understanding irreversible evolution of macroscopic observables from reversible microscopic laws of mechanics, in systems with large number of particles, is an interesting and long-standing question~\cite{Lebowitz_PT1993,Lebowitz_PA1993,lanford1976}. In this context, Boltzmann  made the key observation that  irreversibility is the typical macroscopic behavior given appropriate initial conditions and large number of particles~\cite{boltzmann1897,feynman2017,lanford1976,penrose89,greene2004}. Boltzmann also provided a clear prescription for the construction of an entropy function (which we denote as $S_B$) that is defined for a single microstate of a macroscopic system.  This entropy function is defined for a system in or out of equilibrium, being  equal to the thermodynamic entropy for a system in equilibrium.

The definition of Boltzmann's entropy in a nonequilibrium setting requires one to define the notion of macrovariables and macrostates~\cite{Lebowitz_PA1993,Goldstein_PD2004}. As an example consider  $N$ particles in a box of length $L$. A microstate of the system would be the specification of the position, $\{x_i\}$, and velocities, $\{v_i\}$, of all the particles, $i=1,2,\ldots,N$, which can be represented by a point $X=\{x_i,v_i\}$, in phase space (dentoted by $\Gamma$). A macrovariable is defined as a coarse-grained observable that is a function of the microstate $X$. The value of the macrovariable $M=M(X)$ defines a macrostate $\Gamma_M$ which corresponds to the set of phase space points with the value $M$ of the macrovariable  with a resolution $\Delta M$~\cite{Lebowitz_PA1993,Goldstein_PD2004,Goldstein_Book2020}. One can then associate an entropy to a microstate $X$ as follows. Given a microstate $X$ we first find the value of the macrovariable $M=M(X)$. We then find the volume $|\Gamma_M|$ of the correpsonding macrostate $\Gamma_M$. The Boltzmann entropy, $S_B(X)$, of the microstate $X$ is then defined as
$S_B(X)=k_B \ln |\Gamma_{M(X)}|$ \cite{Goldstein_PD2004,Goldstein_Book2020}. We summarize the above procedure: 
\begin{align}
X \rightarrow M=M(X) \rightarrow |\Gamma_{M(X)}| \to S_B(X)=k_B \ln |\Gamma_{M(X)}|.    
\end{align}
A simple example of a macrovariable is to consider the number of particles in the left and right halves of the box, so in this case $M=(N_L,N_R)$.  For any microstate, $X$, one can count the number of particles in each half of the box and thus determine the value of the macrovariable $M(X)=(N_L(X),N_R(X))$.  Clearly there are many microstates corresponding to the same value $M=(N_L(X),N_R(X))$ for the macrovariable --- their phase space volume $|\Gamma_{(N_L(X),N_R(X))}|$ determines the Boltzmann entropy of the microstate~$X$. 

We note that, for a system not in equilibrium, as the microstate, $X(t)$, evolves with time, the macrovariables $M(t)=(X(t))$ also evolves leading to the evolution of the entropy $S_B(X)$.  In the context of an  interacting dilute gas, Boltzmann argued  that typically a microstate evolves to macrostates with larger and larger volumes which therefore implies an almost monotonic growth of entropy. Eventually the system reaches the macrostate with the largest volume which is in fact the  equilibrium
state.  The monotonic growth of entropy is expected to become a certainty in the limit of large number of degrees of freedom \cite{Lebowitz_PT1993,Lebowitz_PA1993,Lebowitz_Book2008}.  

These ideas involve subtle concepts and it is important to demonstrate and clarify them with specific examples of microscopic models. In particular there are questions on the possible role of ergodicity, mixing and chaos which play important role in thermalization.  There are several  related studies that have investigated the evolution of Boltzmann's entropy previously \cite{AlderWainwright1956,orban1967,Levesque_JSP1993,Rochin_JSP1997,FalcioniPhysicaA2007,Garrido_PRL2004,de2006}.

The free expansion of a gas in a box provides a paradigmatic example to study irreversibility and entropy change in a mechanical model. Several earlier studies~\cite{georgallas1987free,swendsen2008explaining,zanette1991free,bernstein1988expansion,de2017rigourous} have shown that even a non-interacting ideal gas  reaches an equilibrium state if one takes appropriate limit of large number of particles. In particlular Ref.~\cite{de2017rigourous} provide a rigorous proof for the equilibration of the empirical density profile. In an earlier work~\cite{Chakraborti_EM2021},
the evolution of Boltzmann's entropy during free expansion of a classical non-interacting ideal gas was studied for two choices of macrovariables. Two choices of macrovariables and the corresponding  Boltzmann entropies were studied.  The first choice considered a coarse graining of the single particle phase space $\mu \equiv (x,v)$  into rectangles of size $\Delta = \Delta x \Delta v $  and looked at the particle density $f_\alpha=N_\alpha/\Delta$, at time $t$, where $N_\alpha$ is the number of particles in the $\alpha-$th box.  The second choice was the three locally conserved fields $U = \{\rho(x, t), p(x, t), e(x, t)\}$ corresponding to mass density, momentum density and energy density respectively, defined with respect to a spatial coarse-graining length scale $\ell$. The evolution of the   Boltzmann entropies, $S_B^f$ and $S_B^U$,  corresponding to the above two choices  were studied, startting from a single microscopic configuration.  In both case an increase of entropy to the expected equilibrium was observed at long times  implying  irreversible macroscopic evolution. Some interesting observations that were made are the following: (a) For large $N$,  the rate of growth of $S^f_B$, at any given time,  decreases with decreasing  coarse graining scale $\Delta$ and never converges; (b) The relaxation of $S_B^f$ to its equilibrium value is oscillatory with a time period $\tau=2L/\Delta v$; (c) on the other hand, $S_B^U$  shows a monotonic  increase to the equilibrium value and the growth rate converges on decreasing the coarse graining scale  $\ell$; (d) although the equilibrium values of $S^f_B$ and $S^U_B$ match, their values at intermediate times do not match, as is also clear from points (a-c).

The behaviour of $S^f_B$ implied by points (a,b,d) above  are related to the fact that the system considered is non-interacting and integrable. Two crucial features that lead to this behaviour are: (i)  an initial phase space volume $\Delta = \Delta x \Delta v$ will not spread in the velocity  direction, (ii) any two given points with initial  separation,  $(\Delta x, \Delta v)$, will have exact periodic recurrence to the same spatial separation $\Delta x$ with a period $\tau=2L/\Delta v$.  As consequence  of (i) and (ii), fine spatial structures develop in the velocity cell $\Delta v$, with period $\tau$, and these finally lead to the observed features (a,b) of $S^f_B$, as discussed in Ref.~\cite{Chakraborti_EM2021}.  It is expected that the features (i) and (ii) will not  hold in non-integrable systems. In fact we also  expect that feature (ii) may not be valid in interacting integrable systems such as hard rods. Thus it is of great interest to study the evolution of $S^f_B$ and $S^U_B$ in interacting  systems and this is the focus of the present paper.

\begin{figure}
\begin{center}
%\leavevmode
%\includegraphics[width=13.25cm,angle=0]{snap-xv-com.pdf}
%\includegraphics[scale=0.325]{snap-xv-com.pdf}
%\includegraphics[width=13cm,angle=0]{snap-xv.pdf}
\includegraphics[width=14cm]{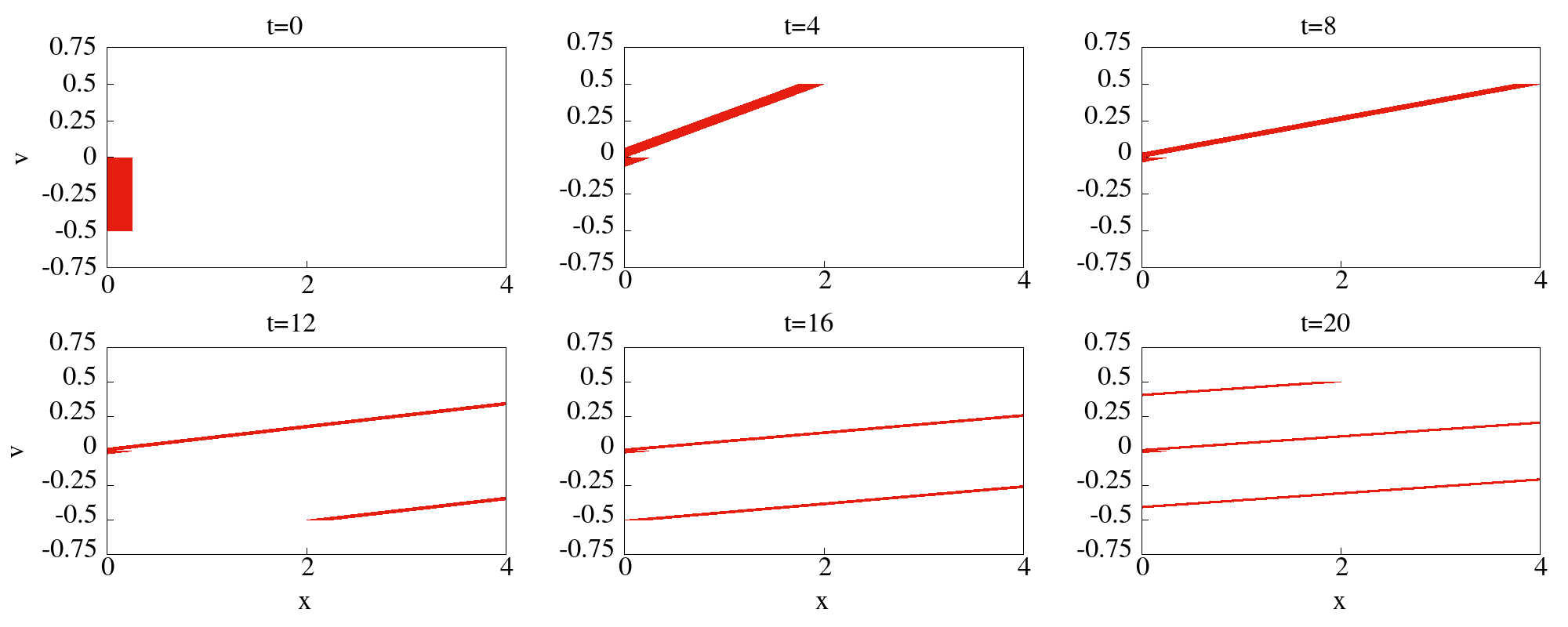}
	\put (-230,170) {$\textbf{{\footnotesize(a) Equal mass gas}}$}

\includegraphics[width=14cm]{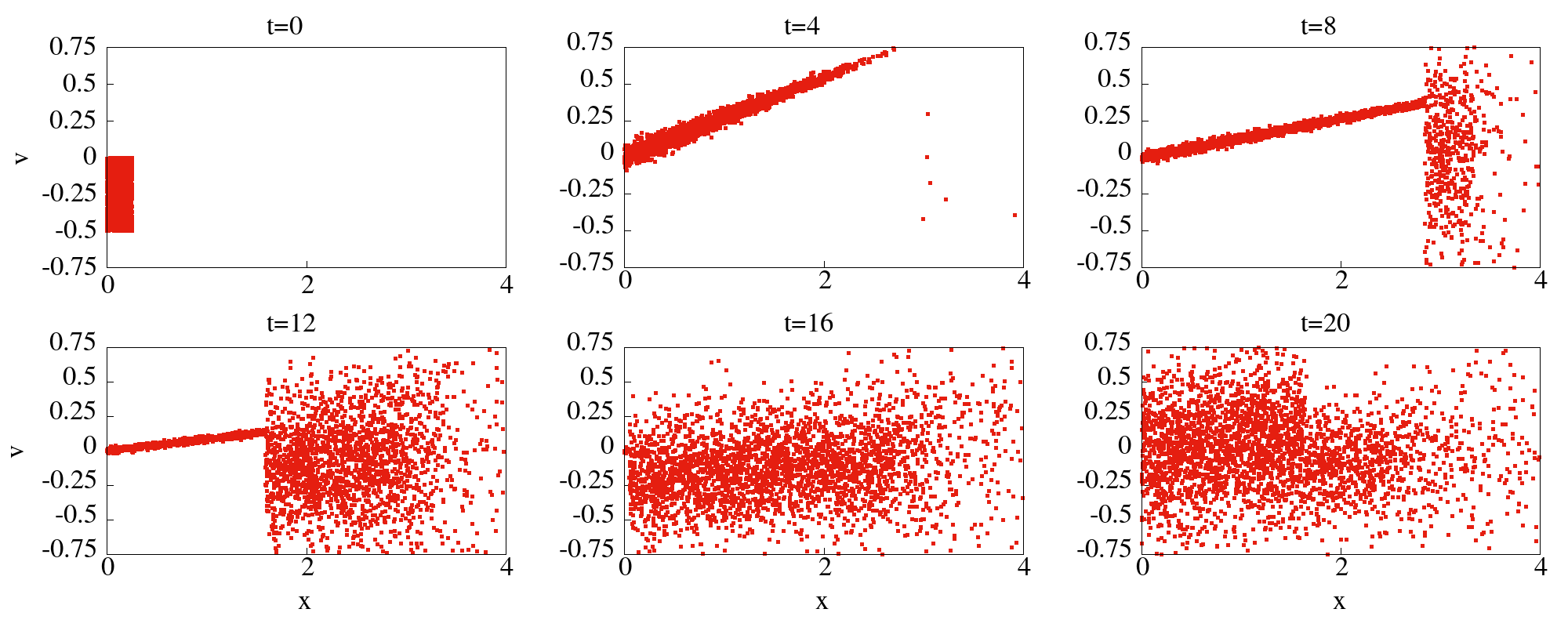}
	\put (-230,170) {$\textbf{{\footnotesize(b) Alternate mass gas}}$}
\caption{Evolution of the distribution of points in single particle phase space for (a) the equal mass hard particle gas and (b) the alternate mass hard particle gas (an example of an interacting system). For both cases, at $t=0$, $N=10000$ particles are uniformly distributed in the cell $x \in (0,0.25), v\in(-0.5,0)$ in a box of length $L=4$ with hard walls at the two ends. In (a) we see a spread only along the spatial direction and fine structures keep developing at all times while in (b) the distribution spreads in both directions and becomes completely diffused at late times. }
\label{fig-xv-com}
\end{center}
\end{figure}

In particular it is of interest to study dilute interacting gases (with short ranged interactions and at low densities) where $f$ and $U$ continue to be good macrovariables~\cite{Goldstein_PD2004}.
Our expectations for such systems are the following: 
\begin{enumerate}
\item The distribution of particles in the single particle phase space   would exhibit  quite different structures for the non-interacting and interacting gases. In the non-interacting case, there is no spread in the velocities and more and more fine structures keep developing in the phase space distribution for all times. New bands keep appearing regularly in the horizontal direction with a period of $2L/\Delta v$. In contrast, for the 
interacting case, the phase space distribution spreads along the velocity direction and quickly becomes diffused in the entire phase space. The difference between the non-interacting and interacting cases can be seen in  the scatter plots [see in Fig.~\ref{fig-xv-com}] of particles evolving in $\mu$-space. Hence, unlike the non-interacting case, here the growth of $S^f_B$ is expected to converge, in the limit $N \to \infty,~\Delta \to 0$, to a monotonically increasing form.

\item In the limit $N\to \infty$ and $\Delta  \to 0$, we expect $\tilde{f}(x_\alpha,v_\alpha,t)=f_\alpha(t)/N$, in the interacting gas,  to satisfy the Boltzmann equation with a collision term:  
\begin{align}
\p_t \tilde{f}(x,v,t)+v\p_x \tilde{f}= [\p_t \tilde{f}]_{\rm coll}.  
\end{align}
For the non-interacting gas, the collision term is absent, and this implies that the entropy $\tilde{s}=\int dx dv \tilde{f} \ln \tilde{f} $ will not grow with time. This is consistent with the observation of vanishing  growth of $S^f_B$ in the limit of $\Delta \to 0$. For the interacting gas we expect the collision term to lead to entropy growth even in the $\Delta \to 0$ limit \cite{lanford1976,Goldstein_PD2004}.

\item In the interacting case we expect local thermal equilibrium (LTE) to be present \cite{spohn2012large}. We set  $\ell=\Delta x$ which is taken to be small but, in the limit of large $N$, each coarse-graining cell still contains a large number of particles. For the dilute gas, the empirical phase space distribution $f_\alpha$ in the $\alpha-$th cell, specified by  $(x_\alpha,v_\alpha)$, should be equivalent to a Gibbs distribution characterised by values of the conserved fields $\rho(x,t), p(x,t), e(x,t)$ in the spatial cell $x=x_\alpha$. For a dilute gas one can also define the velocity and temperature fields: $u(x,t)=p/\rho$ and $T(x,t)=2e/\rho-u^2$. Then, from LTE we expect  the form:
\begin{align}
f_\alpha(x_\alpha,v_\alpha,t)= \frac{\rho(x_\alpha,t)}{\sqrt{2 \pi k_B  T(x_\alpha,t)}}  \exp \left[-\f{m(v_\alpha-u(x_\alpha,t))^2}{2 k_B T(x_\alpha,t)}\right]  
\label{eq:lte}
\end{align}
From this it immediately follows that the two entropies $S_B^f$ and $S_B^U$ will be the same at all times.
\end{enumerate}

A simple model that  incorporates interactions is the alternating mass hard particle gas (AMHPG) which consists of a one dimensional gas of hard point particles  where alternate particles have different masses. The dynamics consists of ballistic motion of the particles in between elastic collisions that conserve both momentum and energy. This system is  non-integrable and known to have good mixing properties even with few particles~\cite{Casati_PRE2003}. A special simplifying feature of this model is that while it is non-integrable, its equilibrium thermodynamic properties are still that of an ideal gas and is known to satisfy standard hydrodynamic evolution \cite{subhadip2021,ganapa2021}.  Hence it is in some sense an ideal candidate which mimics a dilute yet strongly interacting system. The AMHPG has been widely used  in studies concerning the breakdown of Fourier's law of heat conduction in one dimension \cite{Dhar_PRL2001,Grassberger_PRL2002,Casati_PRE2003,Cipriani_PRL2005} and in the verification of the hydrodynamic description~\cite{Garrido_PRL2004,Spohn_JSP2014,Mendl_JSP2016,subhadip2021,ganapa2021,Chakraborti_Splash2022}. Here, we use the AMHPG to investigate  entropy growth during  free expansion  of an interacting gas and to test if the expectations (1-3) above are seen. Our main finding is a validation of this picture.

The paper is organized in the following way. In Sec.~\ref{sec:found} we formulate the basic construction required for the study. This includes the detailed idea of Boltzmann on macroscopic irreversibility leading to definition of entropy, the elaborate description of the model and the choice of macrostates we have used. In Secs.~\ref{sec:choiceI} and \ref{sec:choiceII} we discuss the results obtained from our numerical studies of  the evolution of the macroscopic observables and the corresponding entropy functions for the two different choices of macrovariables. The cases of non-thermal initial conditions and mass ratio close to one are  discussed in Sec.~\ref{NT-IC-mr-1}. In Sec.~\ref{connect-Th} we discuss the connection between Boltzmann entropy $S_B^U$ and Clausius entropy and identify the dissipative current that gives rise to the monotonous growth of entropy. Finally, we conclude with a summary in Sec.~\ref{conclusion}.

%\section{Boltzmann’s entropy, definition of the microscopic model and choice of macrostates}
\section{Definition of model, macrovariables and entropies}
\label{sec:found}

\subsection{Definition of the model}

We consider a one-dimensional interacting gas of $N$ hard point particles with position and velocities given by $x_i$ and $v_i$, $i=1,2,\ldots,N$, confined in a box of length $L$. The only interaction  is through instantaneous elastic collisions between particles which conserve momentum and energy and with the boundary walls. In between collisions, the particles move at constant velocity. We take  particles with $i$ even to have mass $m_1$ and odd indexed ones have mass $m_2 \neq m_1$. When a particle with mass $m_1$ and velocity $v_1$ collides with another particle of mass $m_2$ and velocity $v_2$ then their post-collision velocities are given by
%%%%%%%%%%%%%%%%%%%%%%%%%%%%%%%%%%%%%%%%%%
\begin{align}
\label{elasticcollision1}
v_1^\prime = \frac{ (m_1-m_2) v_1 + 2m_2 v_2 }{m_1+m_2}, \\
v_2^\prime = \frac{2m_1 v_1 + (m_2-m_1) v_2}{m_1+m_2}. \label{elasticcollision2}
\end{align}
%%%%%%%%%%%%%%%%%%%%%%%%%%%%%%%%%%%%%%%%%%
These follow from the conservation of momentum and energy.  At collisions with the walls, at $x=0$ and $x=L$, the boundary particles  are simply reflected. The case $m_1=m_2$ implies that velocities are simply exchanged during collisions and the model effectively maps to the non-interacting case studied in \cite{Chakraborti_EM2021}. 

We consider the initial condition where the $N$ particles are uniformly distributed spatially in the left half of the box while their velocities are chosen from a Maxwell distribution at temperature $T_0$. This corresponds to an equilibrium distribution, in the left half of the box, with temperature $T_0$ and density $\rho_{\rm ini}=2 N/L$. For a given microstate we will consider two choices of macrovariables. We study the evolution of these macrovariables and the associated Boltzmann entropies for the two cases:   (i) starting from a single microstate chosen from the equilibrium distribution (in the left half of the box) and (ii) starting from an ensemble of microstates from the same equilibrium distribution.

\subsection{Definition of macrovariables and entropies}

We describe the two macrovariables that are studied in this work.

{\bf Choice I --- The density of particles in  single-particle phase space:} Here we divide the single particle phase space $\mu \equiv (x,v)$ into cells of size $\Delta_\alpha=\Delta= \Delta x \Delta v$, centered at $(x_\alpha,v_\alpha),~\forall \alpha$. For a given microstate $X=\{x_i, v_i\}$ we count the number of particles, $N_\alpha$, in the $\alpha$-th cell centred at $(x_\alpha, v_\alpha)$.  The empirical phase space density is then given by
%%%%%%%%%%%%%%%%%%%%%%%%%%%%%%%%%%%%%%%%%%
\bea
\label{falpha}
f_\alpha(X)= \frac{N_\alpha(X)}{\Delta_\alpha},
 \eea
%%%%%%%%%%%%%%%%%%%%%%%%%%%%%%%%%%%%%%%%%%
with the normalization $\sum_\alpha f_\alpha(X) \Delta_\alpha =N$. The set of values  $\{f_\alpha(X)=f_\alpha\}$ specifies our  macrovariable corresponding to the microstate $X$. The volume of the $N$-particle phase space  for a macrostate specified by a given set of values of $\{f_\alpha\}$  is   $\Gamma_{\{f_\alpha\}}= \prod_\alpha [ \Delta_\alpha^{N_\alpha}/ N_\alpha !]$, where $N_\alpha=f_\alpha \Delta_\alpha$. Using Stirling's formula in the limit of large $N$, we get, up to an additive constant, the entropy per particle as  
%%%%%%%%%%%%%%%%%%%%%%%%%%%%%%%%%%%%%%%%%%
\bea
s_B^f(X)=S_B^{f}(X)/N=-\frac{1}{N}\sum_\alpha  \Delta_\alpha f_\alpha(X) \ln f_\alpha(X). \label{H-function1}
\eea
%%%%%%%%%%%%%%%%%%%%%%%%%%%%%%%%%%%%%%%%%%
for a given microstate $X$. We have assumed $k_B=1$ throughout the paper. First, we will study the time evolution of this entropy as a single microstate evolves. Second, we will compare this with the entropy obtained by taking an average of the macrovariables over an initial ensemble of microstates. In this context one can define a  single-particle marginal distribution 
%%%%%%%%%%%%%%%%%%%%%%%%%%%%%%%%%%%%%%%%%%
\bea
\label{single_F}
F(x,v,t)=  \sum_{i=1}^N \left\la \delta(x_i(t)-x) \delta(v_i(t)-v) \right\ra,
\eea 
%%%%%%%%%%%%%%%%%%%%%%%%%%%%%%%%%%%%%%%%%%
where the average is  over initial microstates  chosen from the thermal equilibrium ensemble in the left half of the box. To compare with Eq.~\eqref{falpha} we consider  the following distributions with the same  coarse-graining scale $\Delta_\alpha$:
%%%%%%%%%%%%%%%%%%%%%%%%%%%%%%%%%%%%%%%%%%
\bea
F_\alpha(t) = \frac{1}{\Delta_\alpha} \int_{x,v \in \Delta_\alpha} dx ~ dv F(x,v,t). \label{F_alpha}
\eea
We associate to this distribution the following entropy per particle: 
%%%%%%%%%%%%%%%%%%%%%%%%%%%%%%%%%%%%%%%%%%
\bea
s^F_\Delta=-\frac{1}{N}\sum_\alpha  \Delta_\alpha  F_\alpha \ln F_\alpha, \label{s^F}
\eea
Note that this cannot be directly interpreted as a Boltzmann entropy. 
%For a discussion of its interpretation see \redw{appendix \ref{Appa}}. 
Note that for our interacting ideal gas,  $F(x,v,t)$  obeys the Boltzmann-like equation 
\begin{equation}
  \partial_t {F}+v \partial_x {F}=[\partial_t F]_{\rm coll}, \label{Feq}
\end{equation}
where the collision term involves two-particle distributions~\cite{Dhar_PRL2001}.

Typicality, a consequence of law of large number, suggests that as we increase the number of particles in case of a single microstate consistent with the given initial state $\{f_\alpha(0)=F_\alpha(0)\}$, the empirical $f_\alpha(t)$ approaches  the marginal $F_\alpha(t)$ at all times. As a consequence, marginal values of all other fields and observables, such as entropy, will also agree with the respective empirical values in the limit of large $N$. In case of non-interacting gas of equal masses it has been shown that typicality holds \cite{Chakraborti_EM2021}. We expect, that the  interacting gas would also obey typicality which we test in the next section.

{\bf Choice II - The spatial density of mass, momentum and energy:} In this case we divide the box $(0,L)$ into $K=L/\ell$ number of identical subsystems, each having size $\ell$. Now for a particular microstate $X$,  we denote the number of particles, total momentum and total energy in the $a^{\text{th}}$ cell (centered at $x_a=\ell (a-1/2)$ for $a=1,2,...,K$), respectively by $N_a(X)$, $P_a(X)$ and $E_a(X)$. The corresponding macrostate $U$ is specified by the values of the locally conserved quantities $\{N_a,P_a,E_a\}$. 
 Note that for our interacting ideal gas, these are the only conserved quantities and the corresponding fields obey hydrodynamic equations \cite{subhadip2021,ganapa2021}. The densities associated to these conserved quantities are \be
\rho_a=\frac{N_a}{\ell},~ ~p_a=\frac{P_a}{\ell},~e_a=\frac{E_a}{\ell}. 
\label{F_densities}
\ee
From these densities one can define the velocity field,  the internal energy density and temperature field as  
\be
u_a=\frac{p_a}{m\rho_a},~  \epsilon_a=e_a - 
\frac{1}{2}m\rho_a u_a^2, ~
T_a=\frac{\epsilon_a}{\rho_a},
\label{def:u-ep-T}
\ee
with $m=\frac{m_1+m_2}{2}$. 

The volume of the phase-space $|\Gamma_U|$ corresponding to the macrostate $U$ provides the Boltzmann's entropy $S_B^U=\ln |\Gamma_U|$. The volume $|\Gamma_U|$ for given $U$ for ideal gas can be computed easily. In the large $N$ limit, one can express the entropy per particle as
%%%%%%%%%%%%%%%%%%%%%%%%%%%%%%%%%%%%%%%%%%
\bea
\label{Boltzmann_entropy2}
s_B^{U}=\frac{S_B^{^U}}{N}= \frac{1}{N}\sum_a \rho_a \ell  ~s(\rho_a,\epsilon_a),
\eea
%%%%%%%%%%%%%%%%%%%%%%%%%%%%%%%%%%%%%%%%%%
where $s(\rho_a,\epsilon_a)$ is the ideal gas entropy per particle given, up to additive constant terms, by
%%%%%%%%%%%%%%%%%%%%%%%%%%%%%%%%%%%%%%%%%%
\bea
\label{Boltzmann_entropy3}
s(\rho_a,\epsilon_a)= -\ln \rho_a +  \frac{1}{2} \ln \left(\frac{\epsilon_a}{\rho_a} \right)=\ln\left(\frac{\sqrt{T_a}}{\rho_a}\right).
\eea
%%%%%%%%%%%%%%%%%%%%%%%%%%%%%%%%%%%%%%%%%%
 As the microstate $X$ evolves, the macrovariable $U$ evolves and consequently $S_B^U(t)$ also evolves which we study in the next section.

When the initial microstates are chosen from an ensemble corresponding to thermal equilibrium on the left half of the box, we then consider the average of the macrovariables denoted by $\bar{U}=(\bar{N}_a,\bar{P}_a,\bar{E}_a)$  given by
\begin{subequations}
\begin{align}
\bar{N}_a(t) &= \int_{x_a-\ell/2}^{x_a+\ell/2} dx \int_{-\infty}^\infty dv~F(x,v,t), \\
\bar{P}_a(t) &= \int_{x_a-\ell/2}^{x_a+\ell/2} dx \int_{-\infty}^\infty dv~mvF(x,v,t), \\
\bar{E}_a(t) &= \int_{x_a-\ell/2}^{x_a+\ell/2} dx \int_{-\infty}^\infty dv~\frac{1}{2}mv^2F(x,v,t),
\end{align}
\end{subequations}
where $F(x,v,t)$ is defined in Eq.~\eqref{single_F}. Corresponding to the evolution of $\bar{U}$, we study the evolution of $s_B^{\bar U}=S_B^{\bar U}/N$
%%%%%%%%%%%%%%%%%%%%%%%%%%%%%%%%%%%%%%%%%%
\bea
\label{Boltzmann_entropy4}
s_B^{\bar U}= \frac{1}{N}\sum_a \bar{\rho}_a \ell  ~s(\bar{\rho}_a,\bar{\epsilon}_a),
\eea
%%%%%%%%%%%%%%%%%%%%%%%%%%%%%%%%%%%%%%%%%%
where $\bar{\rho}_a=\bar{N}_a/\ell$ and $\bar{\epsilon}_a=[\bar{E}_a-\bar{P}_a^2/(2m\bar{N}_a)]/\ell $.

%\section{Results for the time evolution of macrostates and entropy increase}
\section{Results from simulations}
\label{sec:result}

In this section we present simulation results for the evolution of the macrovariables and the associated entropies   during free expansion of our interacting gas. 
Initially the system is prepared in  thermal equilibrium at temperature $T_0$ in the left half of the box. The particles are uniformly distributed $(0,L/2)$ and the velocities are distributed according to the Maxwell distribution given by
%%%%%%%%%%%%%%%%%%%%%%%%%%%%%%%%%%%%%%%%%%
\be
\label{Maxwell}
g_{\rm eq}(v_i,T_0)=\left( \frac{m_i}{2\pi  T_0} \right)^{1/2} \exp \left(-\frac{m_i v_i^2}{2T_0} \right).
\ee
We set the mass of the odd particles $m_1=2/3$ and of the even particles $m_2=4/3$ such that the effective mass $m=(m_1+m_2)/2=1$. Ideally we should take the limits $N \to \infty, L\to \infty$ keeping fixed density $N/L$ and fixed coarse-graining scale $\Delta x$ so that it contains sufficiently large number of particles. For our system there is no microscopic length scale and the box size $L$ can be scaled out. Hence without loss of generality we have set $L=4$ in all our simulations and consider the limits of large $N$  and $\Delta x \to 0$ (while ensuring large numbers in each cell). Note that temperature can also be trivially scaled out by rescaling time. Hence, again without loss of generality we set $T_0=2.5$.

We perform event-driven molecular dynamics simulation where the positions and velocities of the particles are updated only when a collision event  takes place. Below we show some results for two choices of macrostates starting from either (i) single microstate or (ii) an ensemble of microstates.

\subsection{Choice I of the macrovariables}
\label{sec:choiceI}
\begin{figure*}
\begin{center}
\leavevmode
\includegraphics[width=14cm,angle=0]{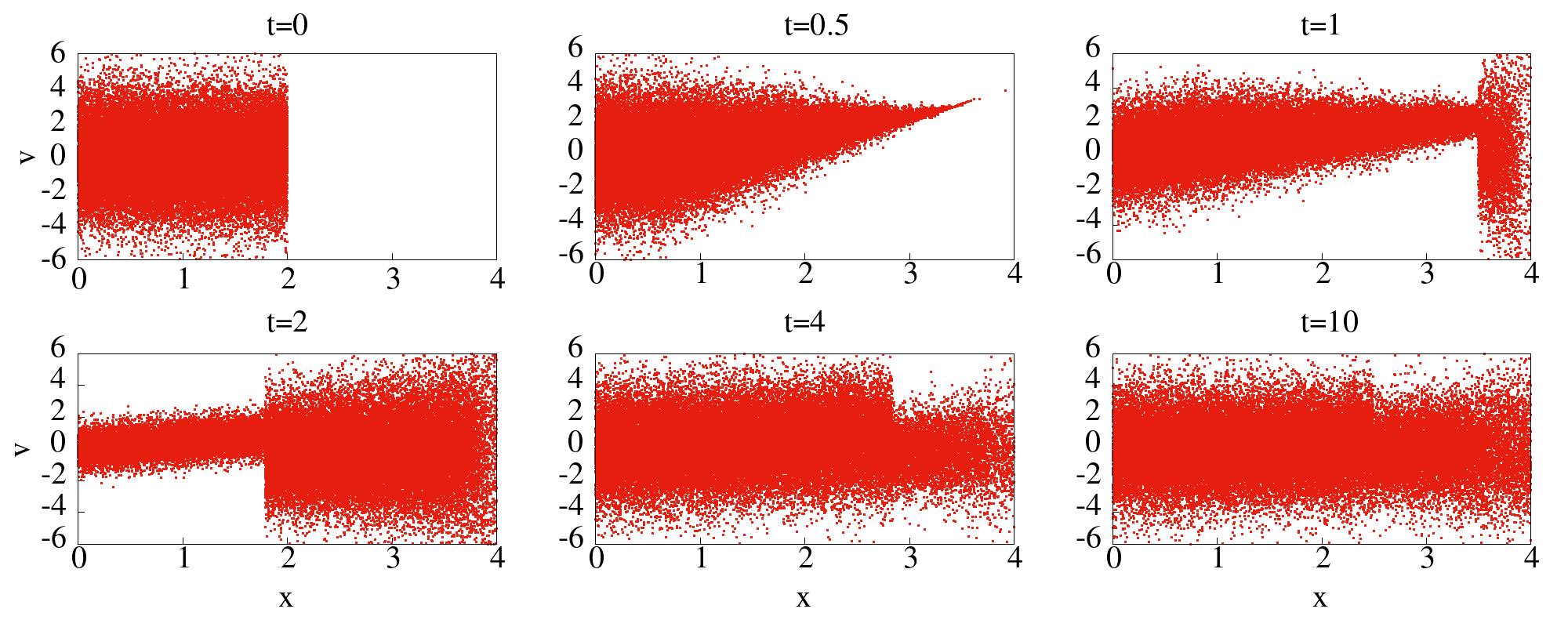}
\caption{Distribution of $N=50000$ particles in $x$-$v$ plane at different times $t=0,~0.5,~1,~2,~4$ and $10$. Initially all the particles with alternate masses $m_1=2/3,~m_2=4/3$ are distributed uniformly within $(0,L/2)$ with $L=4$. The initial velocities are chosen from Maxwell distribution given by Eq.~\eqref{Maxwell} with temperature $T_0=2.5$. We notice a moving shock front which becomes more prominent after the collision with the right wall. The speed of the shock front  is roughly associated with the fastest moving particles and can be seen to decrease with time because of interactions. }
\label{fig-xv}
\end{center}
\end{figure*}

We first consider a random single realization of the system consisting of $N=50000$ particles chosen from the above mentioned canonical distribution in the left half of the box. In Fig.~\ref{fig-xv} we show the evolution of the  distribution of particles in the $x$-$v$ space. Unlike the equal mass case, we observe spreading of the distribution both in $x$ and $v$ directions due to re-distributon of velocities through elastic collisions. We notice a shock front which is formed by the fastest particles. This front becomes more prominent after the collision with the right wall and speed of the shock front decrease with time because the overall speed of the fastest moving particles and their number at the front  decrease due to interactions. At late time the distribution reaches a new equilibrium in the full box. 
%in a non-trivial way until the new equilibrium is reached unlike in equal-mass case where the mixing in space direction was very fast and repeated structure was developed in the velocity direction. 

\begin{figure}
\begin{center}
\leavevmode
\includegraphics[width=14cm,angle=0]{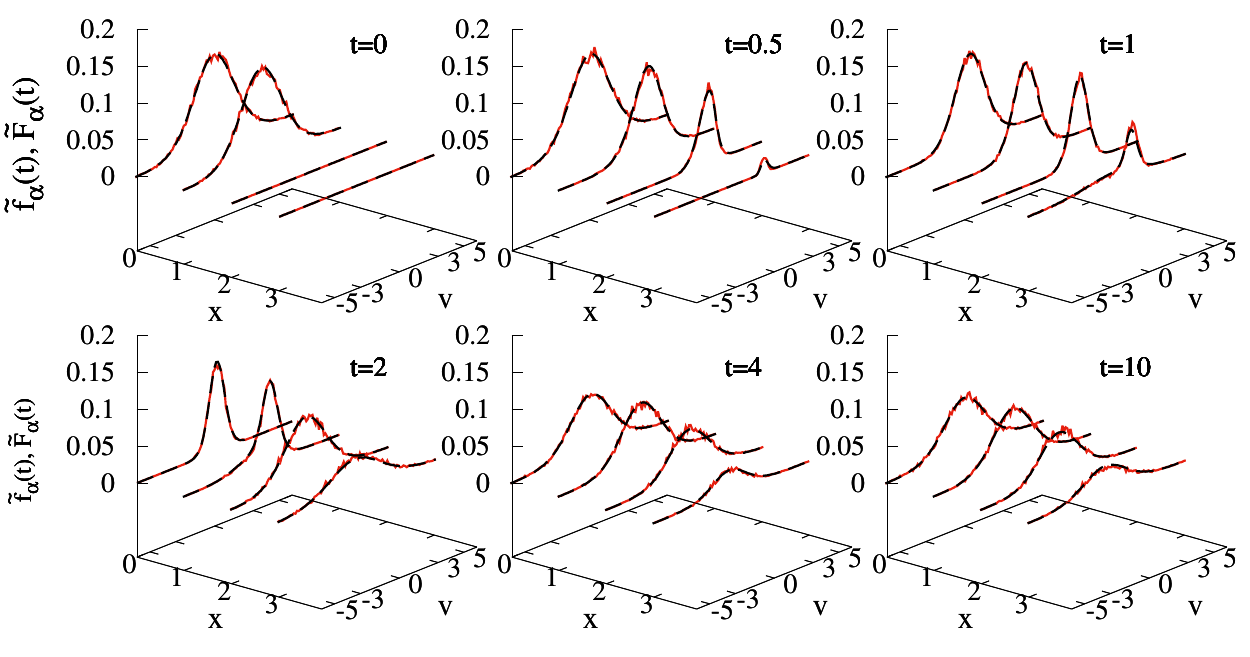}
\caption{Plot of the single particle distribution $f_\alpha(x,v,t)$ as given by Eq.~\eqref{falpha} at different times. The initial condition is the same as in Fig.~\ref{fig-xv}. The red solid lines are empirical $\tilde{f}_\alpha(x,v,t)=f_\alpha(x,v,t)/N$. The black dashed lines are the corresponding marginals $\tilde{F}_\alpha(x,v,t)=F_\alpha(x,v,t)/N$ [defined in Eqs.~\eqref{single_F},~\eqref{F_alpha}] calculated by taking average over $10^4$ realizations of identical systems with particle number $N=4000$. We see a very good agreement between the empirical and marginal distribution functions which indicates the validity of typicality in this interacting system. The grid size was taken as $\Delta x=1$ and $\Delta v=0.125$.}
\label{fig-f_alpha}
\end{center}
\end{figure}

\begin{figure}
\begin{center}
\leavevmode
\includegraphics[width=14cm,angle=0]{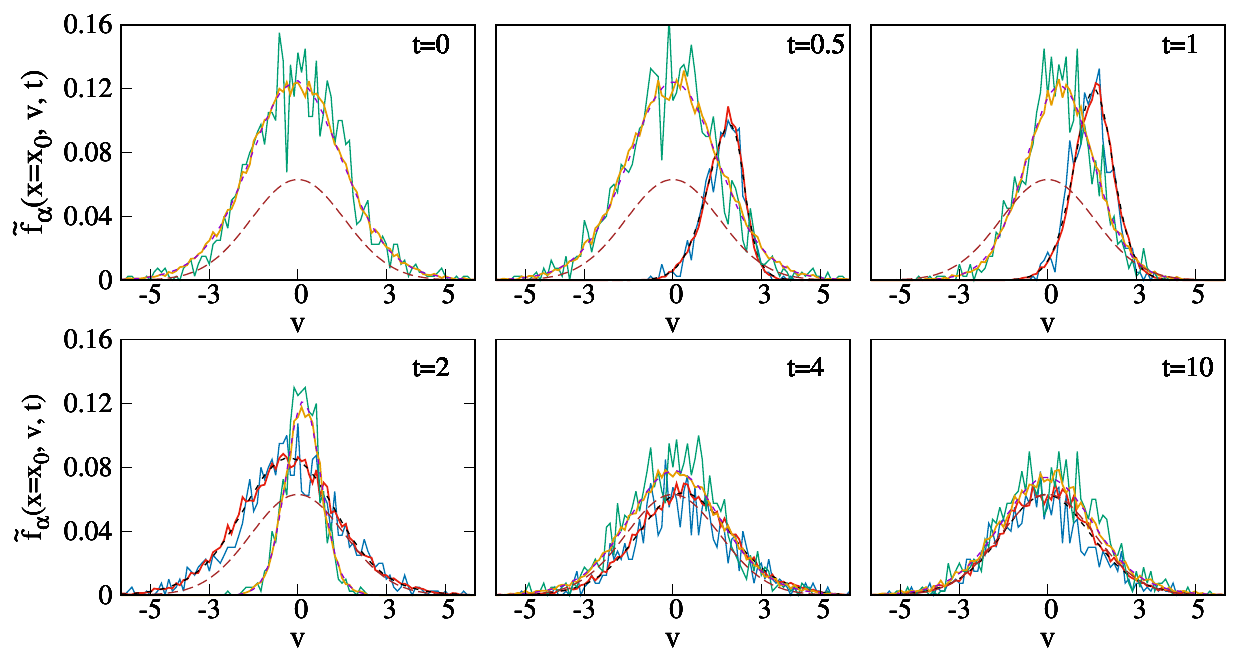}
\caption{ Here we plot the velocity distribution $f_\alpha(x,v,t)$  in two spatial cells at $x=L/4$ (first cell) and $x=3L/4$ (third cell), at different instants of times where we divide the system into four spatial cells as in Fig.~\ref{fig-f_alpha}. The initial condition is also the same in Fig.~\ref{fig-xv}. For spatial cell at $x=L/4$, we plot the empirical $\tilde{f}_\alpha(x,v,t)=f_\alpha(x,v,t)/N$ for $N=5000$ (green thin solid lines) and $N=50000$ (yellow thick solid lines). The magenta dashed lines are the corresponding marginals $\tilde{F}_\alpha(x,v,t)=F_\alpha(x,v,t)/N$ calculated by taking average over $10^4$ realizations of identical systems with particle number $N=4000$. Similarly, for the cell at $x=3L/4$, we plot $\tilde{f}_\alpha(x,v,t)$ for $N=5000$ (blue thin solid lines) and $N=50000$ (red thick solid lines). The black dashed lines are the respective marginal values. Note that at $t=0$, there is no profile for $x=3L/4$ cell as there is no particle at that moment. In addition to that, we plot the final equilibrium distribution by brown dashed lines in each panel. We observe that the fluctuations in the empirical fields decrease with increasing $N$ and they approach the mean profiles thus demonstrating  the validity of typicality. The profiles for both cells approach the equilibrium one at long times.}
\label{fig-f_alpha2}
\end{center}
\end{figure}

To compute the single particle empirical distribution $f_\alpha(x,v,t)$ defined in Eq.~\eqref{falpha} we divide  the $\mu$-space ($x$-$v$ space) into equal grids of size $\Delta_\alpha = \Delta x \Delta v$ and count the number of particles in each cell. For the  choice $\Delta x = 1$ and $\Delta v = 0.125$, 
we  show in Fig.~\ref{fig-f_alpha} the profile of $\tilde{f}_\alpha=f_\alpha/N$, at different instants of time. These are given by the  solid red lines. 
We also compute the  mean profile $\tilde{F}_\alpha=F_\alpha/N$, where $F_\alpha$ is given by Eq.~\eqref{F_alpha},  by considering an average over $10^4$ realizations from a microcanonical ensemble of the system with  energy $E_0=NT_0/2$. The black dashed lines in Fig.~\ref{fig-f_alpha} represent $\tilde{F}_\alpha$. For the single realization we considered $N=5\times 10^4$ particles while for the ensemble average we considered  number $N=4\times10^3$.   We notice that both the empirical $\tilde{f}_\alpha$ and marginal $\tilde{F}_\alpha$ agree quite well.   In Fig.~\eqref{fig-f_alpha2} we show a more detailed comparison between $\tilde{f}$ and $\tilde{F}$ in the cells at $x=L/4$ and $x=3L/4$, at different times, and see very good agreement between the two.
A few interesting qualitative features that we notice are: (i) At $t=0$ out of the four spatial cells, two are empty while two are uniformly filled with particles  having Maxwell velocity distribution; (ii) As time evolves the fast particles move into the third and fourth cells and consequently we see velocity distributions developing in those cells. These distributions are not normalized since the area under the curves correspond to the mean densities in each cell which change with time and eventually reach the equilibrium value; (iii) In Fig.~\eqref{fig-f_alpha2}  we see that at times $t=0.5$ and $t=1.0$, there  is a positive mean velocity in the third cell but the gas is cooler  and the distribution is asymmetric.  At time $t=2.0$, the gas in this cell is hotter than in the first cell; (iv) In Fig.~\eqref{fig-f_alpha2}  we also see clearly the approach to the final equilibrium state. The agreement between $\tilde{f}_\alpha$ and  $\tilde{F}_\alpha$  signifies  typicality in our interacting system, in the sense that the mean macroscopic behaviour and that of individual trajectories are the same.   


\begin{figure}
\begin{center}
\includegraphics[width=7cm,angle=0]{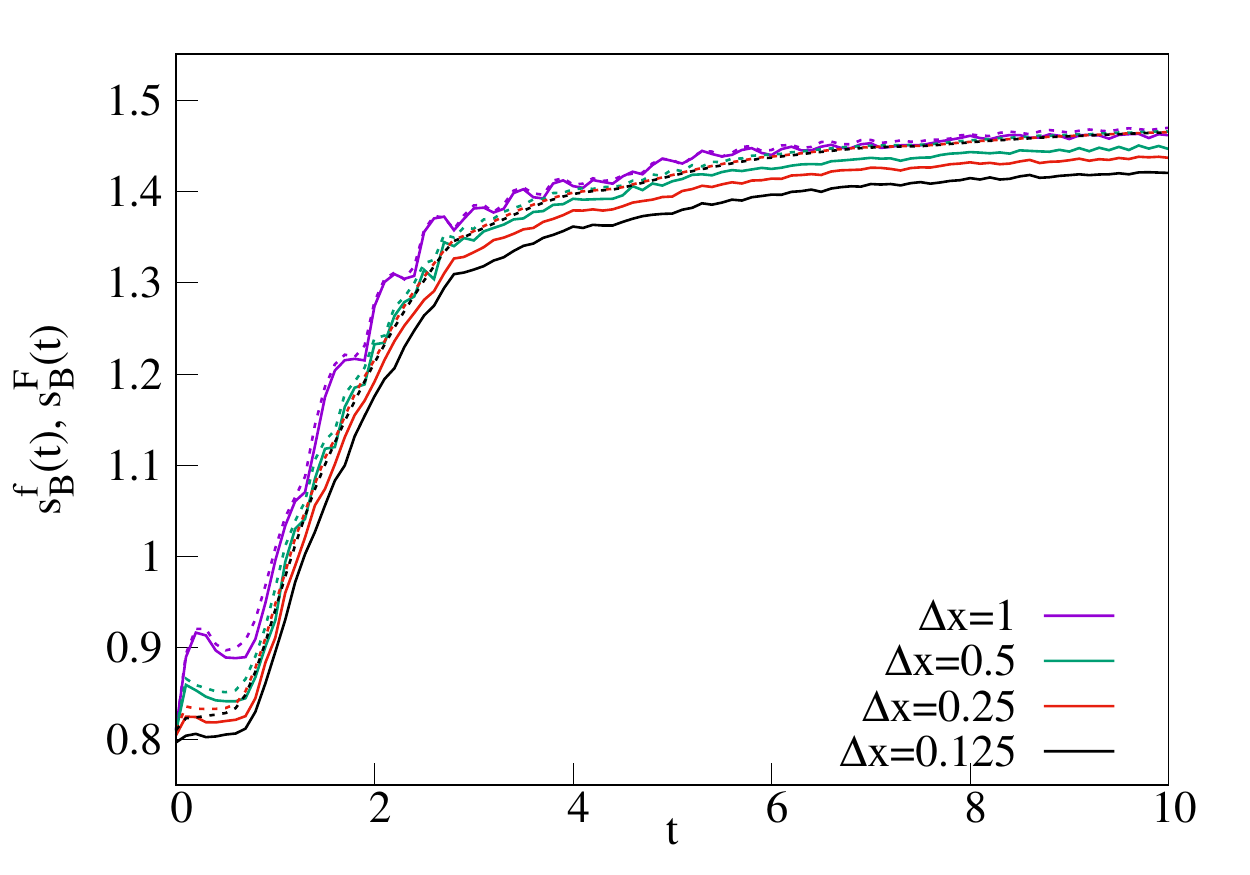}
	\put (-100,120) {$\textbf{{\footnotesize(a)}}$}
\includegraphics[width=7cm,angle=0]{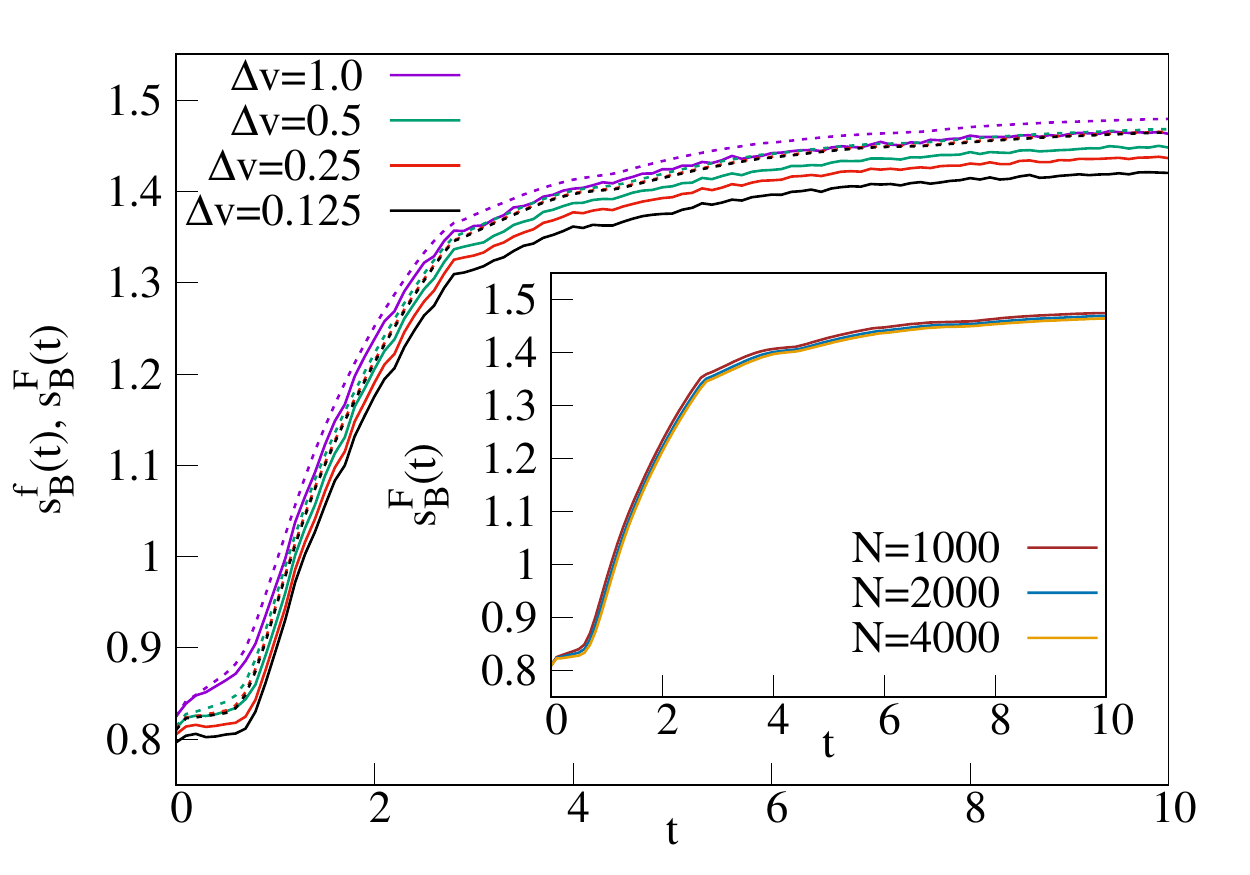}
	\put (-100,120) {$\textbf{{\footnotesize(b)}}$}
\caption{Plot showing $s^f_B(t)$ (solid lines) for $N=50000$ and $s^F_B(t)$ (dashed lines) for for $N=4000$ with time for different $\Delta$. The initial condition and the parameter values are the same as used in Fig.~\ref{fig-f_alpha}. Panel (a): We plot for different $\Delta x$ while the velocity grids remain unchanged at $\Delta v=0.125$. Panel (b): We vary $\Delta v$ and keep $\Delta x = 0.125$ fixed. In both cases we see the entropies increase monotonically. We see convergence of the entropy growth curve, $s^F_B(t)$ as we decrease the grid size. However, $S^f_B(t)$ appears to converge initially but then starts decreasing, due to finite size effects (see text). In the inset of panel (b) we plot marginal $s^F_B(t)$ for a fixed grid size $\Delta x=0.125$ and $\Delta v=0.125$ but for different particle numbers $N=1000, 2000$ and $4000$. We notice the marginal $s^F_B(t)$ has been converged.}
\label{fig-sF}
\end{center}
\end{figure}

We show in Fig.~\eqref{fig-sF}a the evolution of the entropies  $s^f_B(t)$ (solid lines) from  Eq.~\eqref{H-function1} and $s^F_B(t)$ (dashed lines)  from Eq.~\eqref{s^F}, for (a) different values of $\Delta x$ with fixed $\Delta v$ and (b) different values of $\Delta v$ with fixed $\Delta x$.  In all cases $s^f_B$ was evaluated from a single trajectory with $N=5 \times 10^4$ particles while  $s^F_B$ was evaluated for $N=4 \times 10^3$  particles by averaging over $10^4$ realizations. The additive constants in the entropy expressions are fixed in such a way that at $t=0$, $s^F_B(0)$ agrees with $s^U_B(0)$ [given by Eq.~\eqref{Boltzmann_entropy4}].  We observe a non-monotonous (oscillatory) increase of both $s^f_B$ and $s^F_B$ when $\Delta x$ is large, however, this non-monotonicity disappears as we go to lower $\Delta x$. This oscillatory behaviour, as will be explained later, is related to the movement of the shock front (see Fig.~\ref{fig-xv}) across spatial cells of size $\Delta x$ ( see Sec.~\ref{sec:choiceII}). 
 In panel (b) we observe  monotonous increase of entropy with time for all $\Delta v$. In  both the panels we notice that, with decreasing cell size $\Delta$, the entropy curves,  $s^F_B$, converges to a single curve. However, $s_B^f$ initially appears to converge on decreasing $\Delta$, but fails to do so on further decease in cell size $\Delta$ which is a consequence of finite size effects. On the other hand, as shown by dashed lines in Fig.~\ref{fig-sF}, the entropy $s^F_B$ for the marginal distribution $F$ converges with decreasing $\Delta$. Moreover,  in the inset of Fig.~\ref{fig-sF}b we also observe convergence with increasing $N$ for fixed $\Delta$.

\begin{figure}
\begin{center}
\includegraphics[width=7.5cm,angle=0]{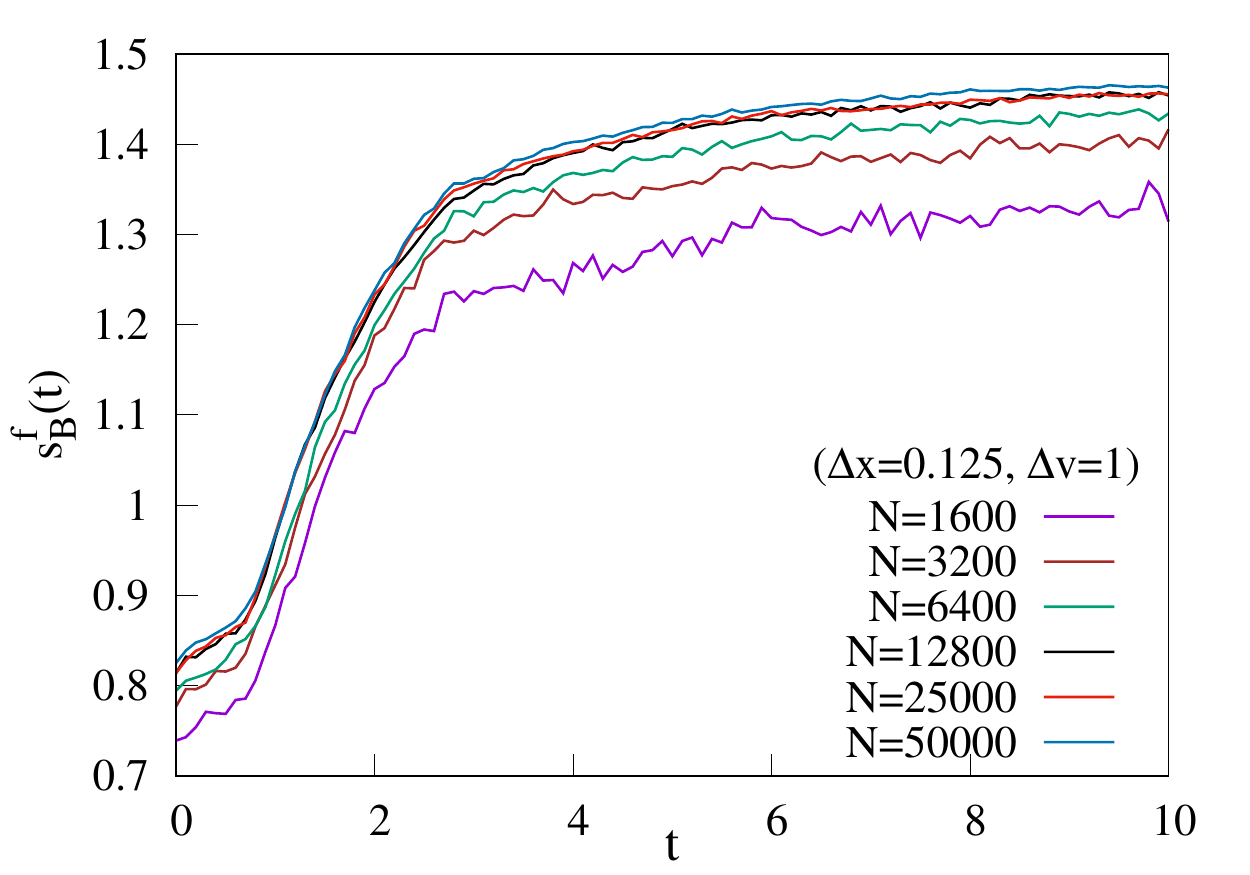}
	\put (-180,130) {$\textbf{{\footnotesize(a)}}$}
\includegraphics[width=7.5cm,angle=0]{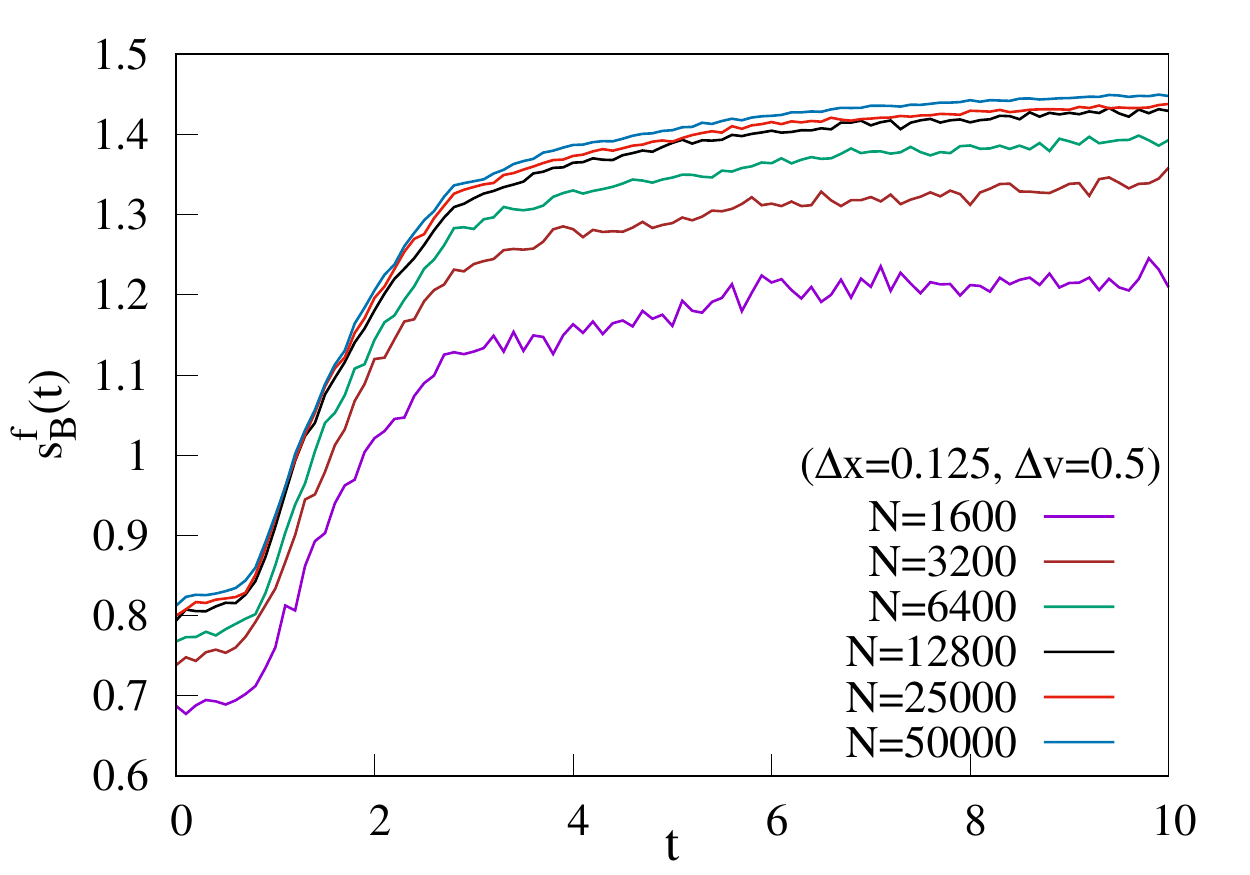}
	\put (-180,130) {$\textbf{{\footnotesize(b)}}$} \\
\includegraphics[width=7.5cm,angle=0]{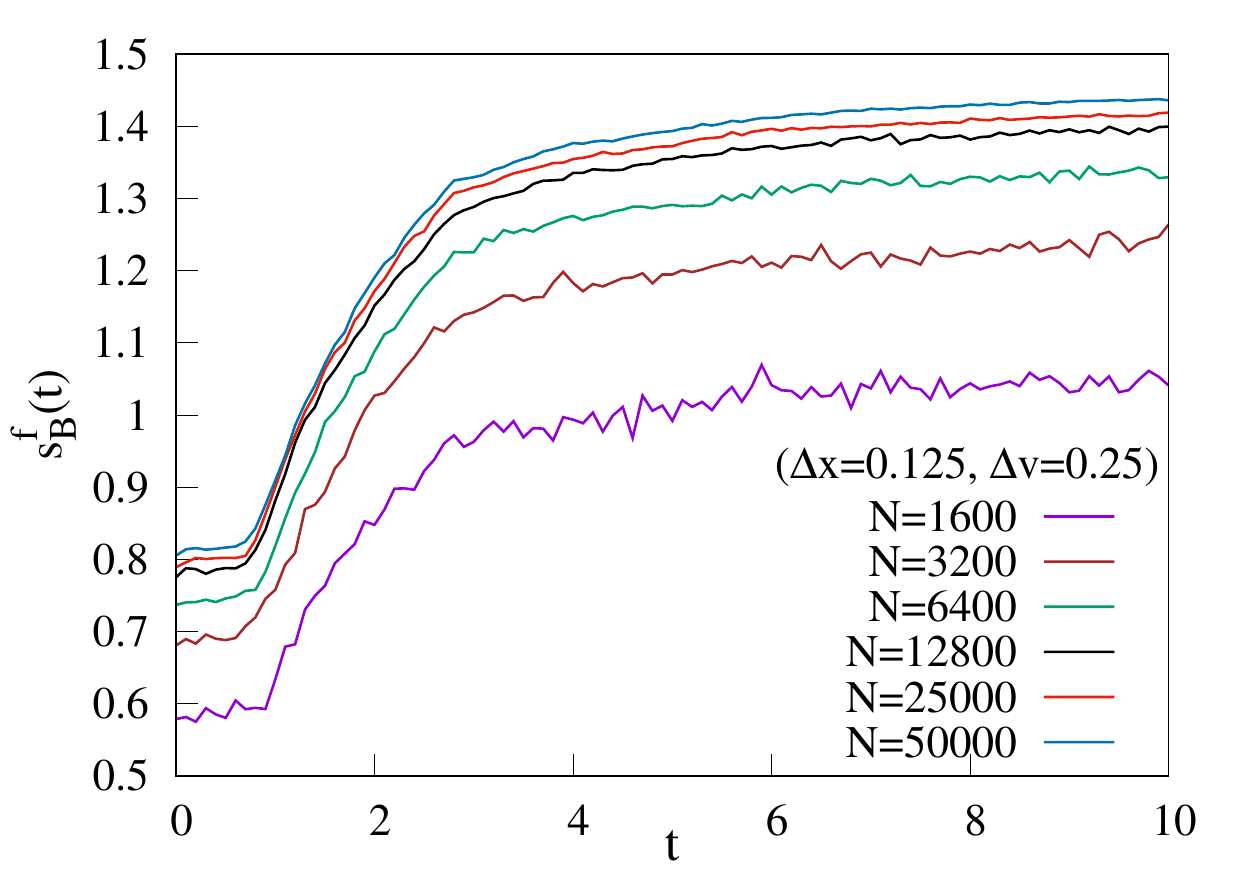}
	\put (-180,130) {$\textbf{{\footnotesize(c)}}$}
	\includegraphics[width=7.5cm,angle=0]{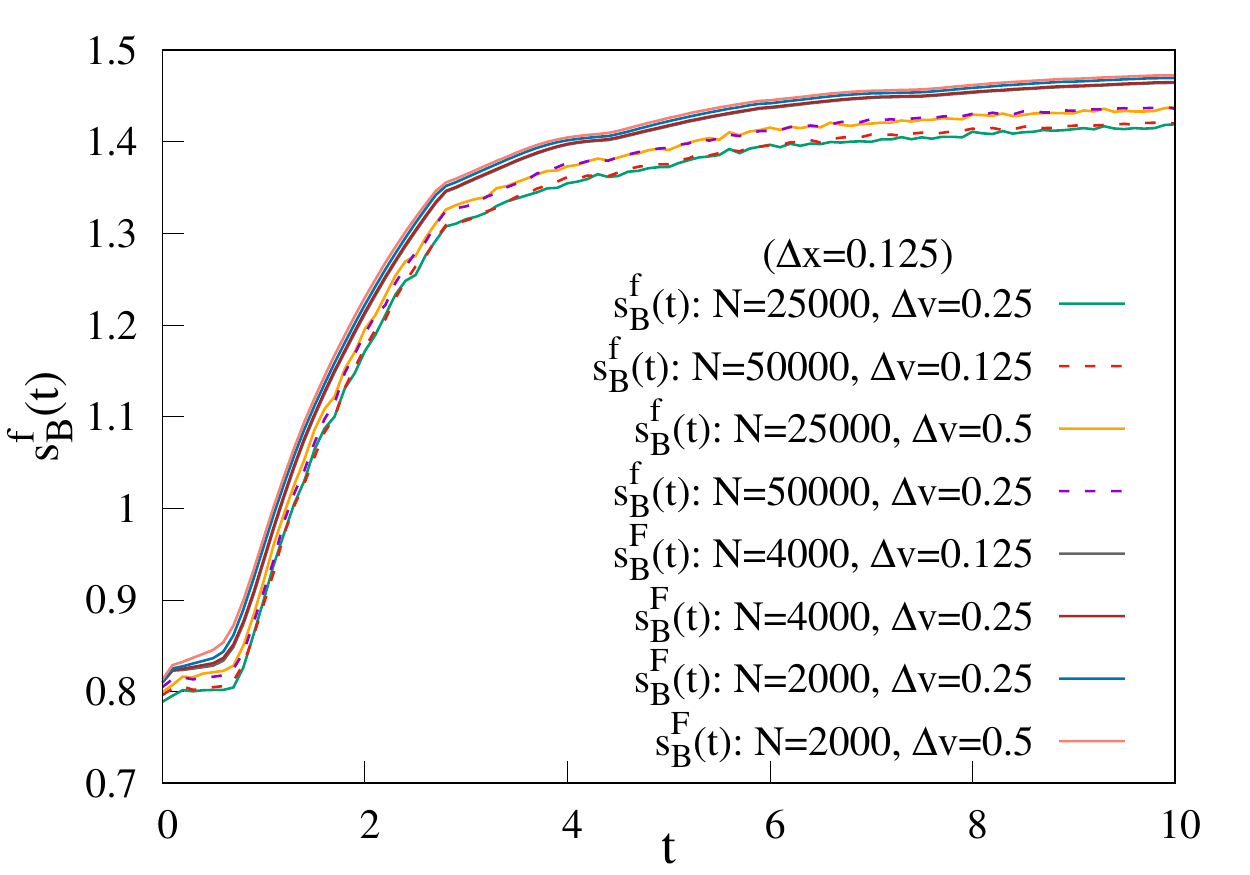}
	\put (-180,130) {$\textbf{{\footnotesize(d)}}$}
\caption{Plot showing evolution of $S^f_B(t)$ with time and their dependece on particle number $N$ and  coarse-graining size $\Delta v$. We keep fixed $\Delta x =0.125$ and show the convergence of entropy with increasing system size for (a)$\Delta v=1.0$, (b)$\Delta v=0.5$ and (c) $\Delta v=0.25$. In (d) we see collapse of the $s^f_B$ evolution data for different values of $N$ and $\Delta $ keeping $N_\Delta = \frac{N\Delta}{Lv_{th}} $ fixed at two values, $\sim 123$ and $246$, respectively. We observe as $N_\Delta$ increases the $s^f_B$ converges to the expected limiting curve.  This limiting curve is obtained from the data collapse of $s^F_B$ with different $N$ and $\Delta$ keeping $N_\Delta$ fixed (at values $\sim 10$ and $20$) and convergence among the collapsed curves for different $N_\Delta$ (solid lines). 
The initial condition and the parameter values are the same as used in Fig.~\ref{fig-f_alpha}. }
\label{fig-sFp}
\end{center}
\end{figure}

We now explain the finite size effects observed in Fig.~\ref{fig-sF}.  Ideally, one would expect a convergence of the entropy growth curves provided the limit $N \to \infty$ is taken before the $\Delta \to 0$ limit while keeping $N_\Delta= \frac{N \Delta}{Lv_{th}}$ fixed at some  large value. Here $v_{th} \sim \sqrt{T}$ is the typical speed. We illustrate this behavior in Fig.~\ref{fig-sFp} where in panels (a), (b), (c), we plot the evolution of $s_B^f$ for $\Delta v=1.0,~0.5$ and $0.25$, respectively, with $\Delta x=0.125$. In all cases we observe convergence with increasing $N$ for fixed $\Delta =\Delta x \Delta v$. In Fig.~\ref{fig-sFp}d we observe the expected collapse of the entropy growth curve for different values of $N$ and $\Delta$ keeping $N_\Delta$ fixed. However a collapsed curve corresponding to a given $N_\Delta$ is not the  limiting  growth curve of $s_B^f$, which would be obtained asymptotically in the limit of large $N_\Delta$ and is harder to achieve numerically.   In Fig.~(\ref{fig-sFp}d), we plot 
the collapsed curves for both $s^f_B$ and $s^F_B$, for two different sets of values of $N_\Delta$ (different for $s^f_B$ and $s_B^F$).  We see that for $s^F_B$ we already see convergence giving us the  asymptotic $s_B^F$ curve. On the other hand we see that with increasing $N_\Delta$ the collapsed $s^f_B$ curves approach the asymptotic $s_B^F$ curve. We thus note that the asymptotic   growth curve of $s_B^F$ (expected to be the same as for $s_B^f$) is obtained at relatively smaller system sizes compared to $s_B^F$. 
 
% or the marginal entropy $s^F_B$ this approach is faster compared to the empirical entropy $s^f_B$.
 %converges to a monotonously increasing graph as we decrease grid size where the final increase of entropy is obtained to be equal to $\ln(2)$. On the other hand, the empirical entropy $s^f_B$ is converging more slowly. 

\begin{figure*}
\begin{center}
\leavevmode
\includegraphics[width=7.5cm,angle=0]{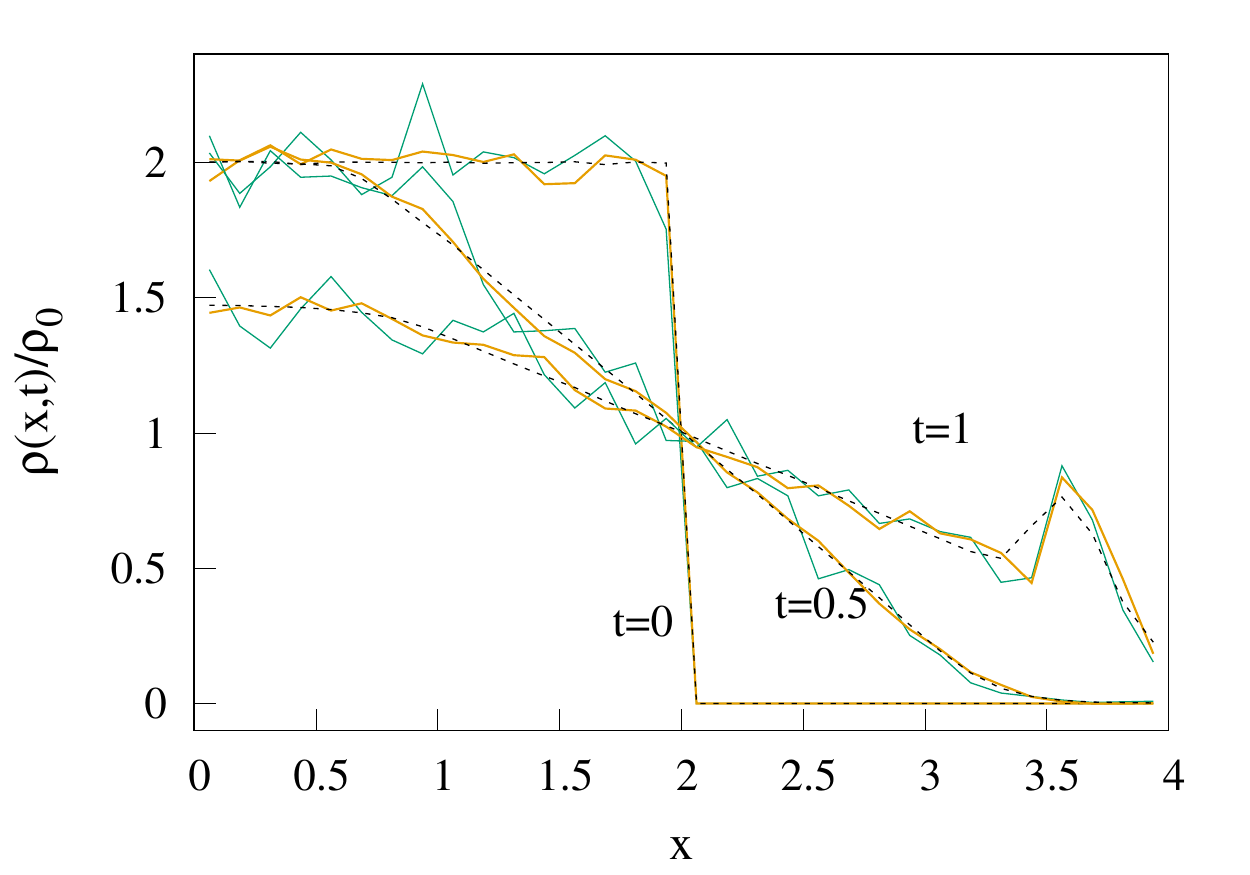}
\includegraphics[width=7.5cm,angle=0]{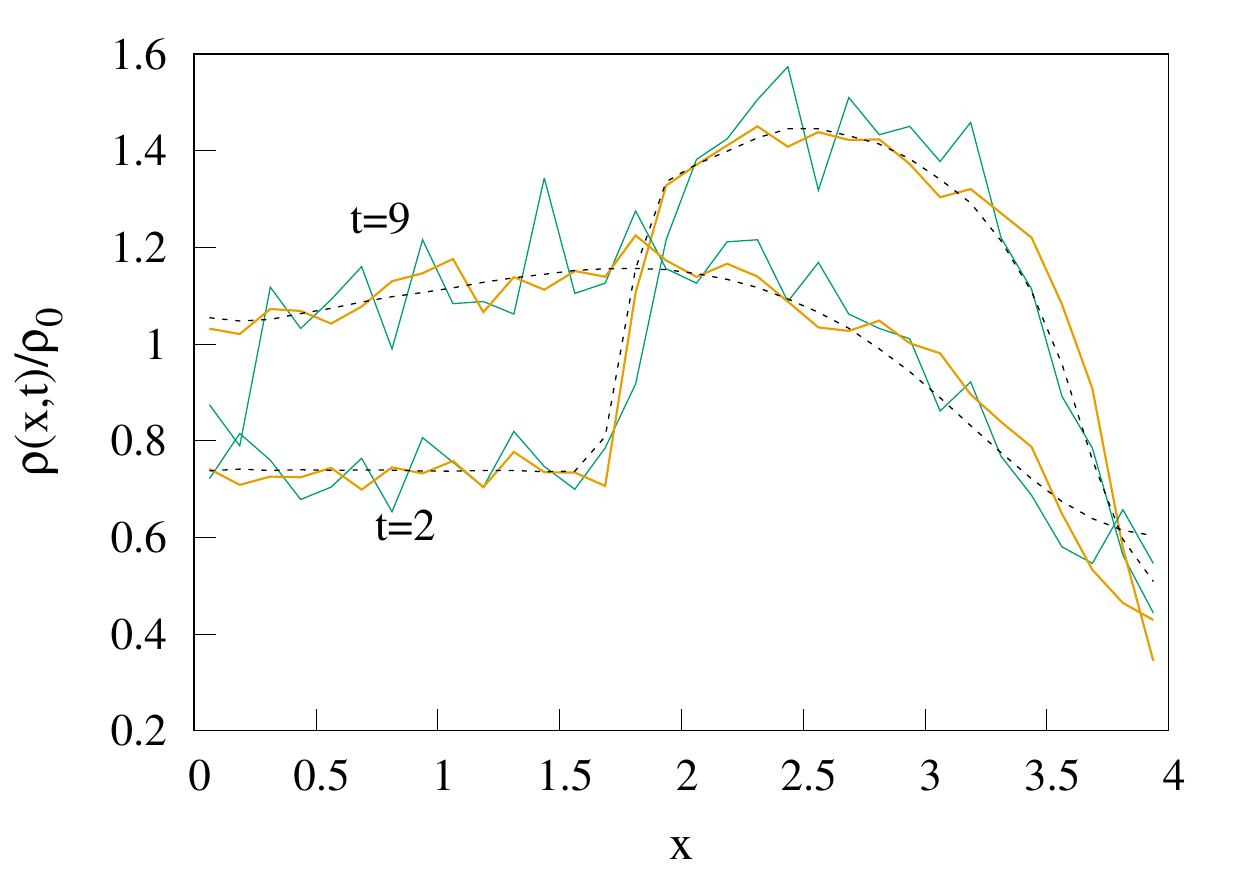}
\includegraphics[width=7.5cm,angle=0]{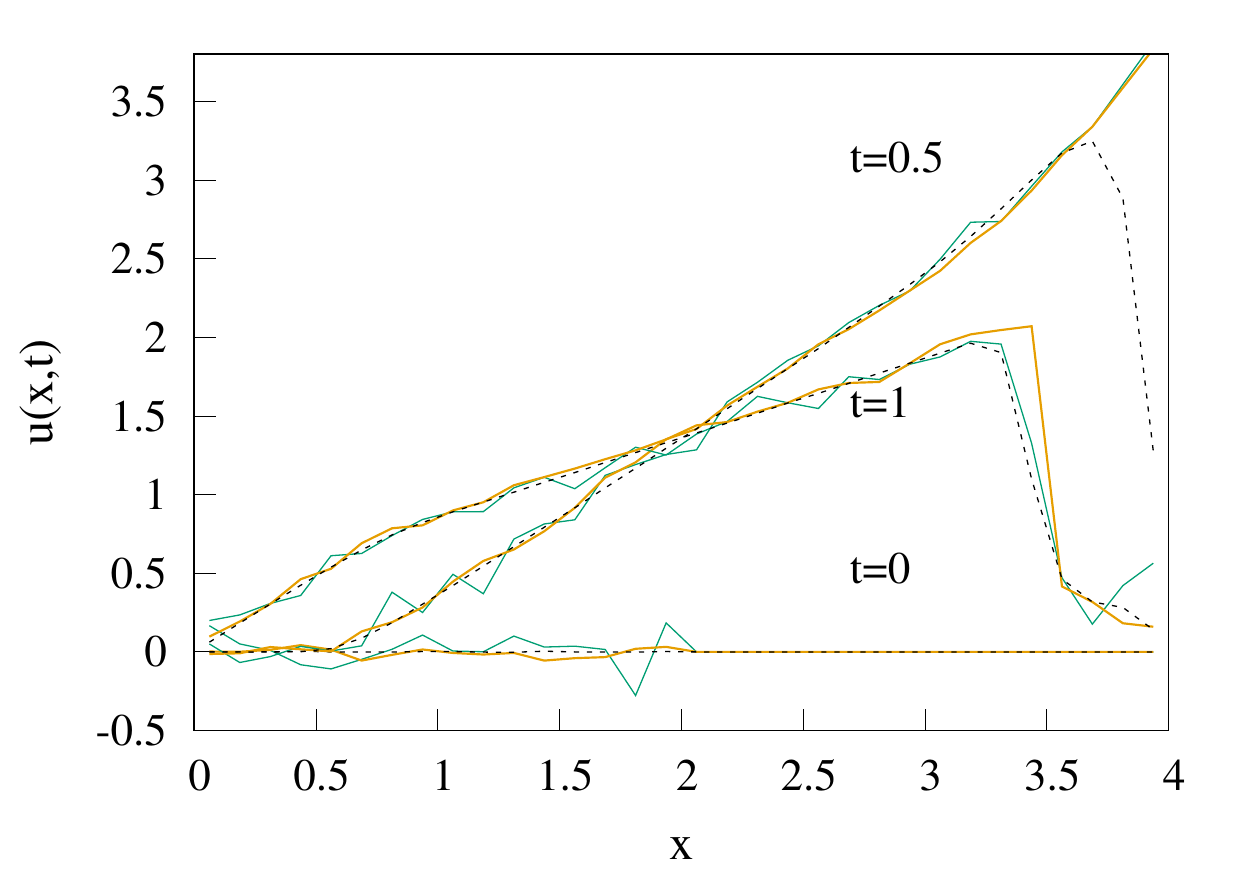}
\includegraphics[width=7.5cm,angle=0]{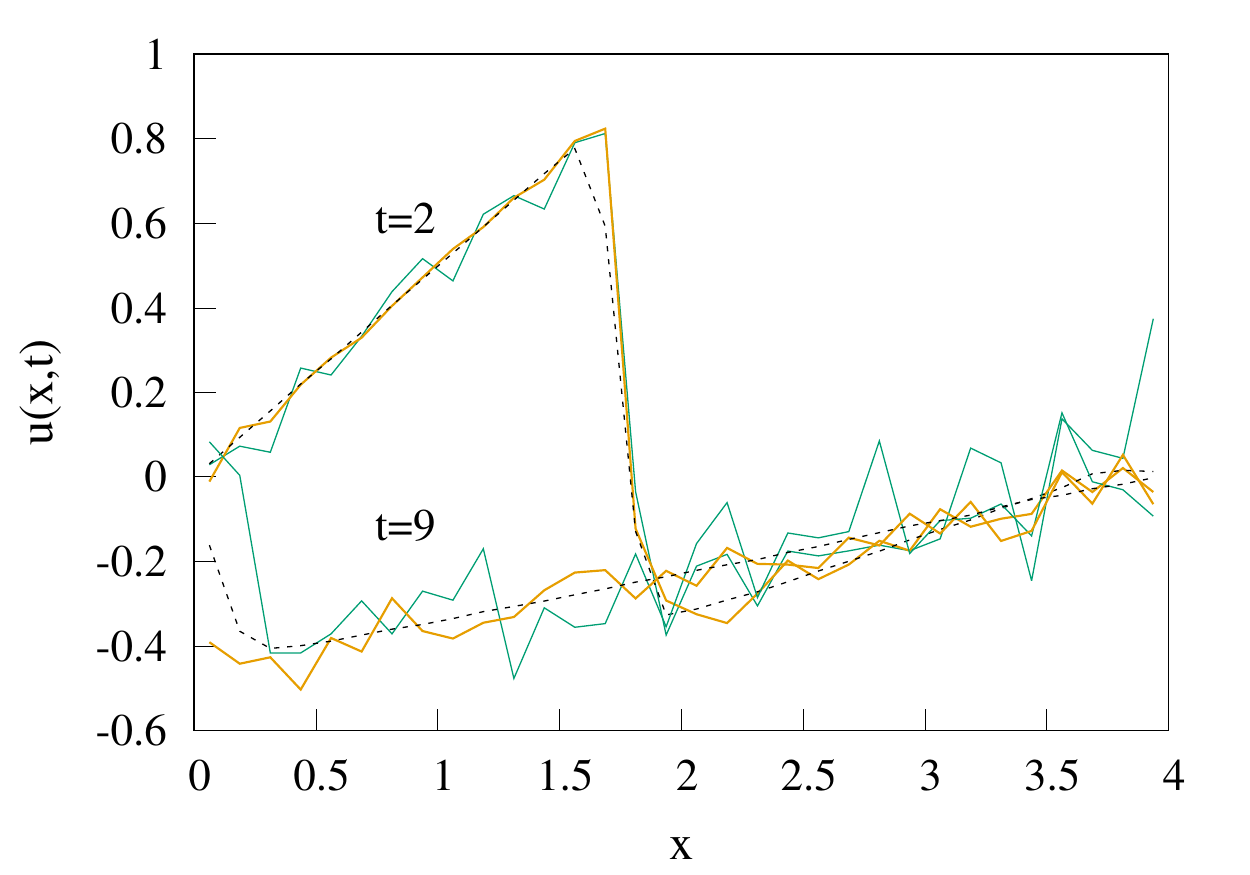}
\includegraphics[width=7.5cm,angle=0]{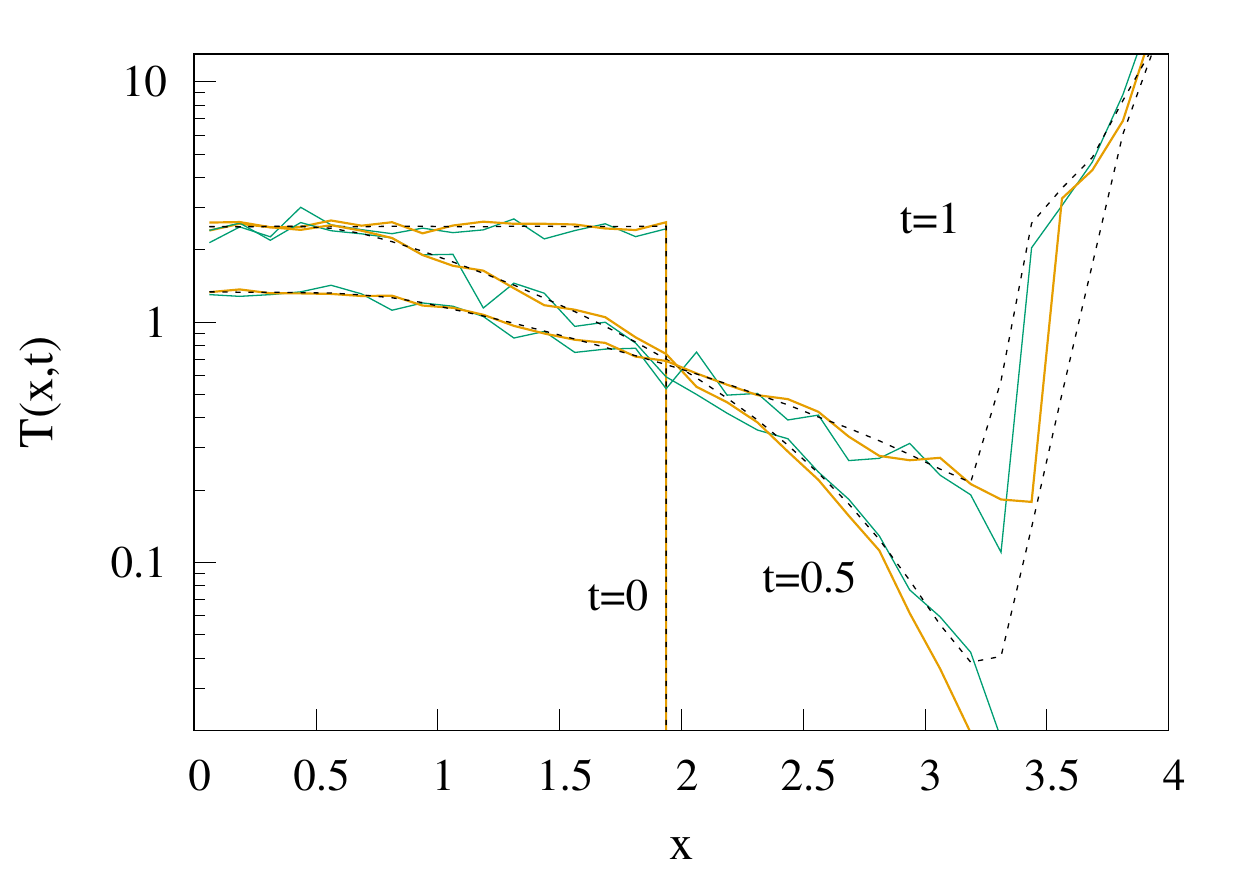}
\includegraphics[width=7.5cm,angle=0]{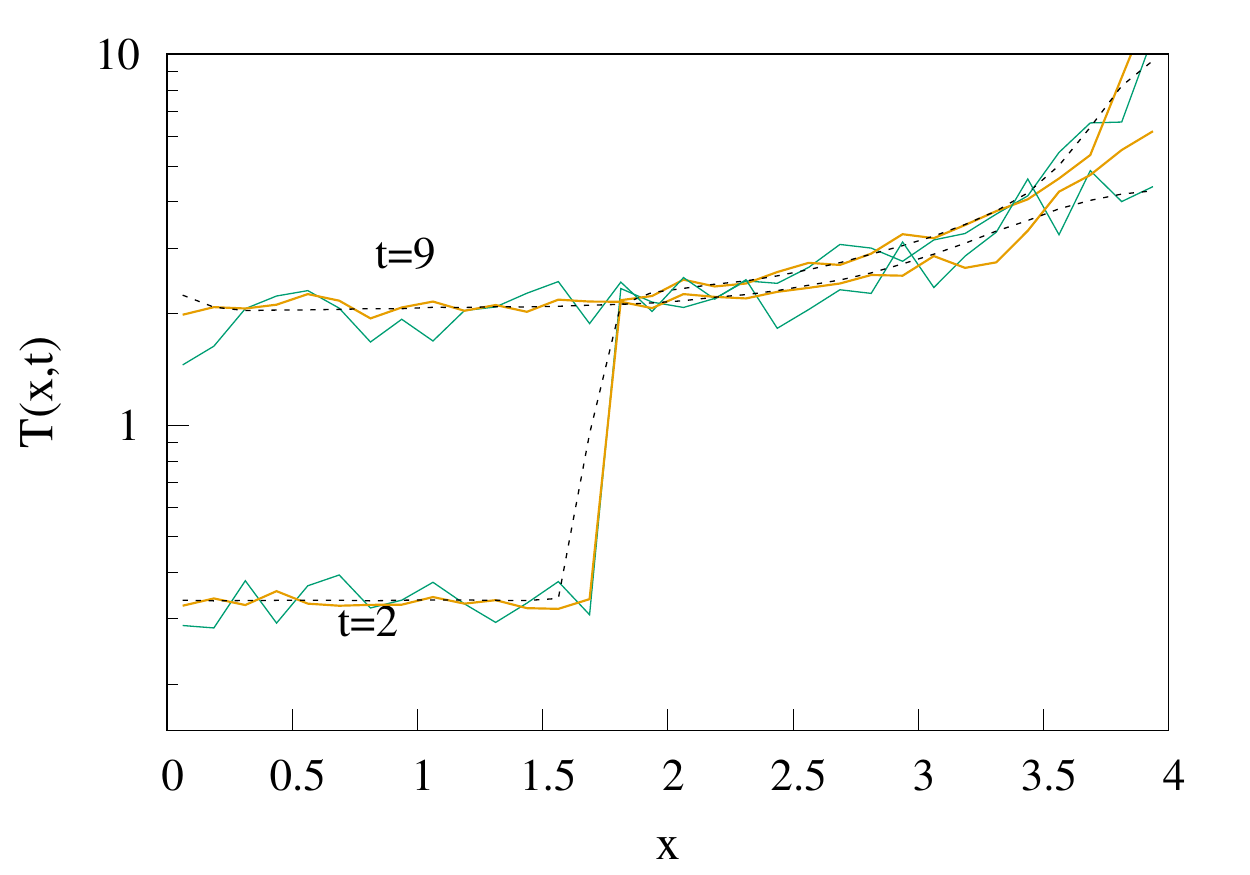}
\caption{Plot of the evolution of spatial profiles (solid lines) of three locally conserved fields density ${\rho}(x,t)$, velocity ${u}(x,t)$, and temperature ${T}(x,t)$  together with the corresponding marginal profiles (dashed lines) $\bar{\rho}$, $\bar{u}$, and $\bar{T}$ obtained from the initial condition and parameter values given in Fig.~\ref{fig-xv} and with $\ell=0.125$. The density is normalized by the mean value $\rho_0=N/L$. Left and right panels shows profiles at respectively smaller and higher times.  Empirical profiles are plotted for $N=5000$ (green solid lines) and $50000$ (yellow solid lines) whereas the marginal profiles for $N=4000$ were  calculated by averaging over $10^4$ initial configurations. We see that the fluctuations in the empirical fields die out with increasing $N$ and they approach towards the corresponding mean profiles, thus verifying typicality. We notice a shock in the profiles at times $t=1$ and $t=2$, just after the expanding gas hits the right wall. At $t=9$, the profiles are approaching the final flat equilibrium profiles.  
}
\label{fig-typicality}
\end{center}
\end{figure*}

\begin{figure*}
\begin{center}
\leavevmode
\includegraphics[width=5cm,angle=0]{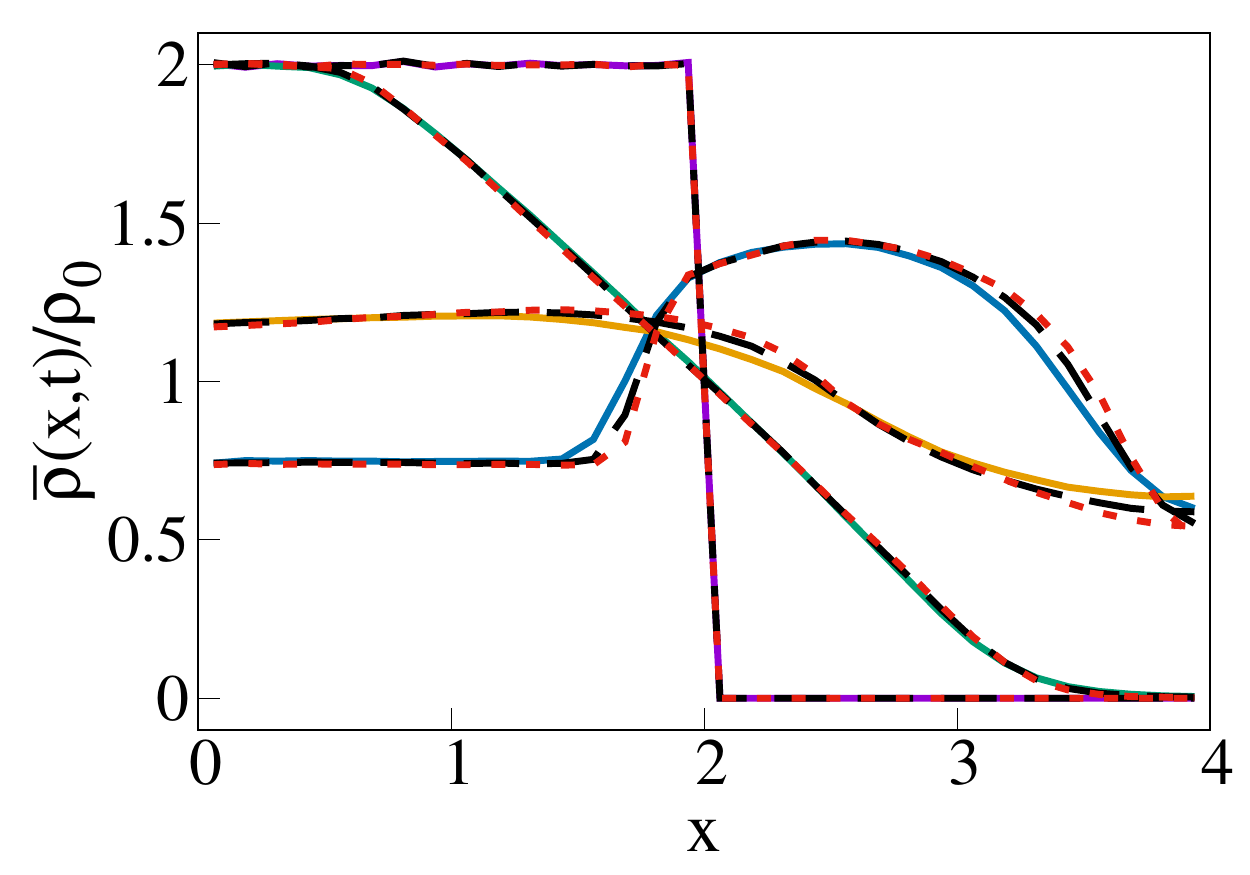}
\includegraphics[width=5cm,angle=0]{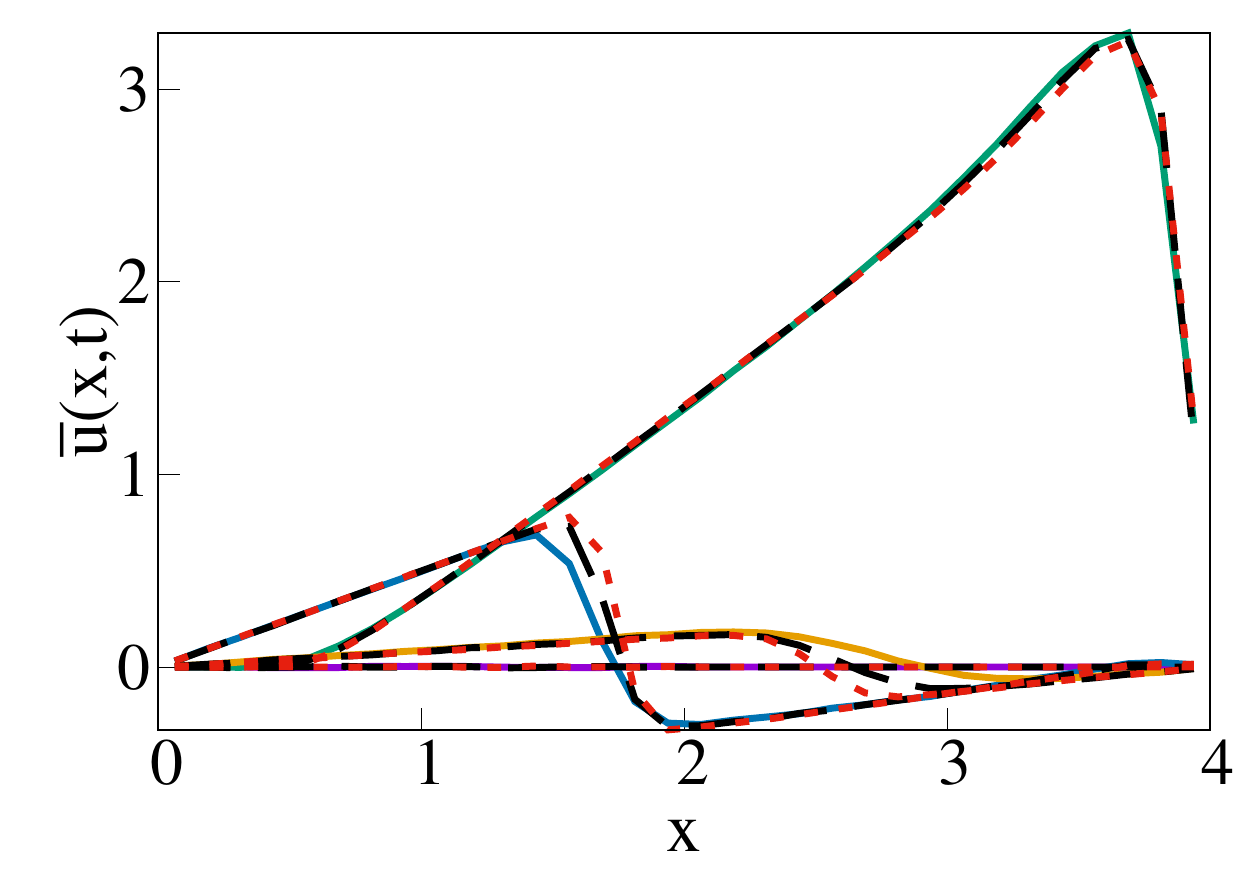}
\includegraphics[width=5cm,angle=0]{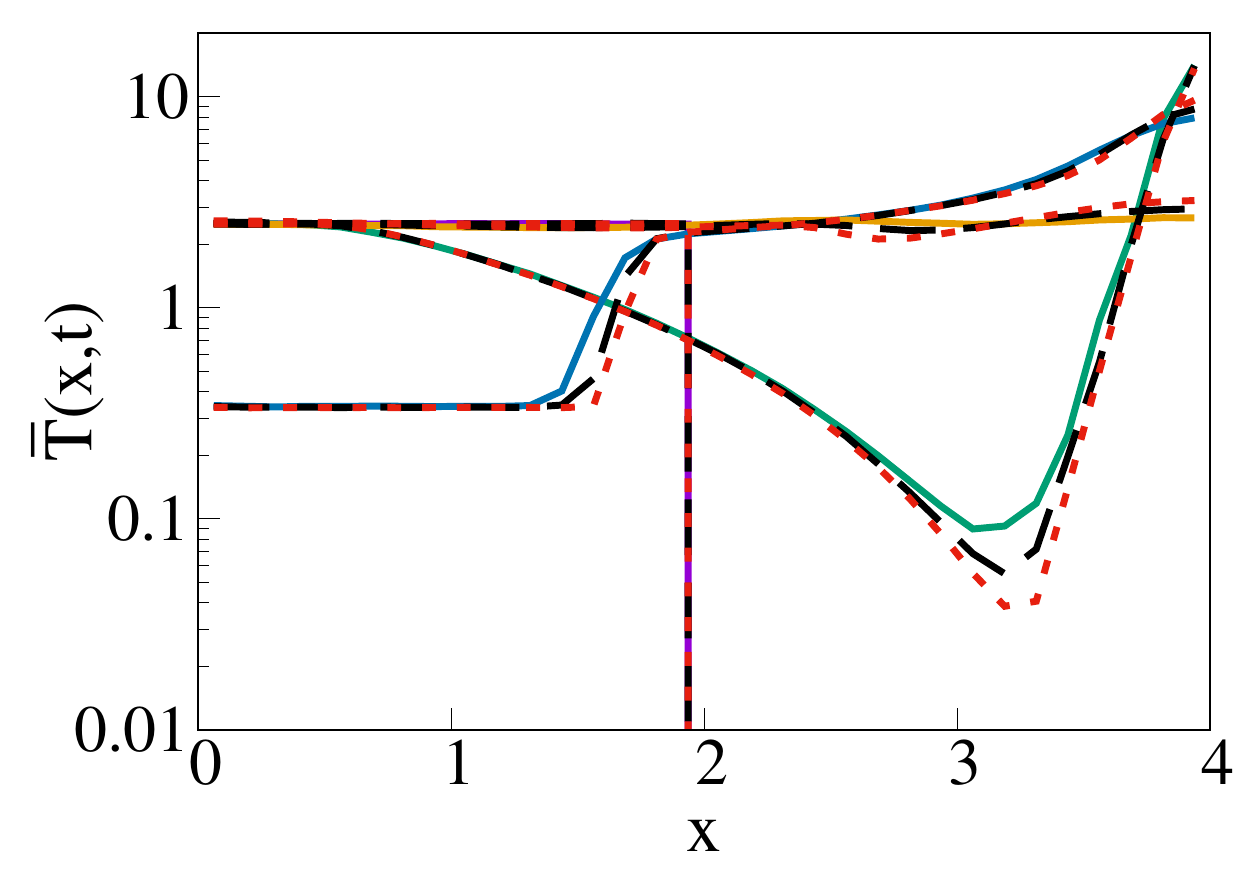}
\caption{Plot of the marginal spatial profiles  $\bar{\rho}(x,t)$,  $\bar{u}(x,t)$, and  $\bar{T}(x,t)$  at different times $t=0$ (magenta lines), $0.5$ (green lines), $2$ (blue lines) and $10$ (yellow lines) for $N=1000$. The initial condition and parameter values are same as  in Fig.~\ref{fig-xv}. The density is normalized by the mean value $\rho_0=N/L$. To check the convergence with $N$, we plot all three profiles at the above mentioned times for $N=2000$ (black dashed lines) and $N=4000$ (red dotted lines). The excellent agreement  assures  that $N=4000$ is sufficiently large enough to perform marginal calculations.}
\label{fig-fields-marginal}
\end{center}
\end{figure*}

\newpage

\subsection{Choice II of the macrovariables}
\label{sec:choiceII}
To compute our second set of macrovariables, $U=(N_a,P_a,E_a)$, we divide the box $(0,L)$ into $K$ equal cells each of length $\ell=L/K$ centred at $x_a=(a-{1}/{2})\ell$ for $a=1,2,...,K$ and evaluate the fields $\{\rho_a,u_a,T_a\}$ defined in Eqs.~\eqref{F_densities} and \eqref{def:u-ep-T}. As for choice I, we again study the evolution of macrovariables starting from either a single microstate  or  from an ensemble of microstates corresponding to a thermal ensemble in the left half of the box.

In Fig.~\ref{fig-typicality}  we plot the empirical fields obtained from a single trajectory for particle numbers $N=5000$ and $5\times 10^4$.  We also plot the marginal fields  computed by taking average over $10^4$ trajectories  with $N=4000$ particles and total energy $E_0=NT_0/2$. We observe that the fluctuation of empirical fields around the mean marginal value decreases with increasing particle number and they converge to the corresponding marginal profiles, expected as a consequence of typicality. In principle these profiles can be understood from the solution of Euler equations till the time the gas hits the right boundary. Interestingly, after the gas gets reflected from the right wall, a visible discontinuity appears due the motion of the shock in all the three profiles as can be seen from the plots at $t=1$ and $t=2$ in  Fig.~\ref{fig-typicality}. The positions of the shocks are the same  those observed in Fig.~\ref{fig-xv}.

We further check whether $N=4000$ is sufficient to provide  profiles of the marginal fields. In order to do that we plot the marginal profiles, namely, density $\bar{\rho}(x,t)$, velocity $\bar{u}(x,t)$ and temperature $\bar{T}(x,t)$ in Fig.~\ref{fig-fields-marginal} for different particle numbers $N=1000$, $2000$ and $4000$. We find that the convergence is good enough at $N=4000$.

\begin{figure}
\begin{center}
\leavevmode
\includegraphics[width=7.5cm,angle=0]{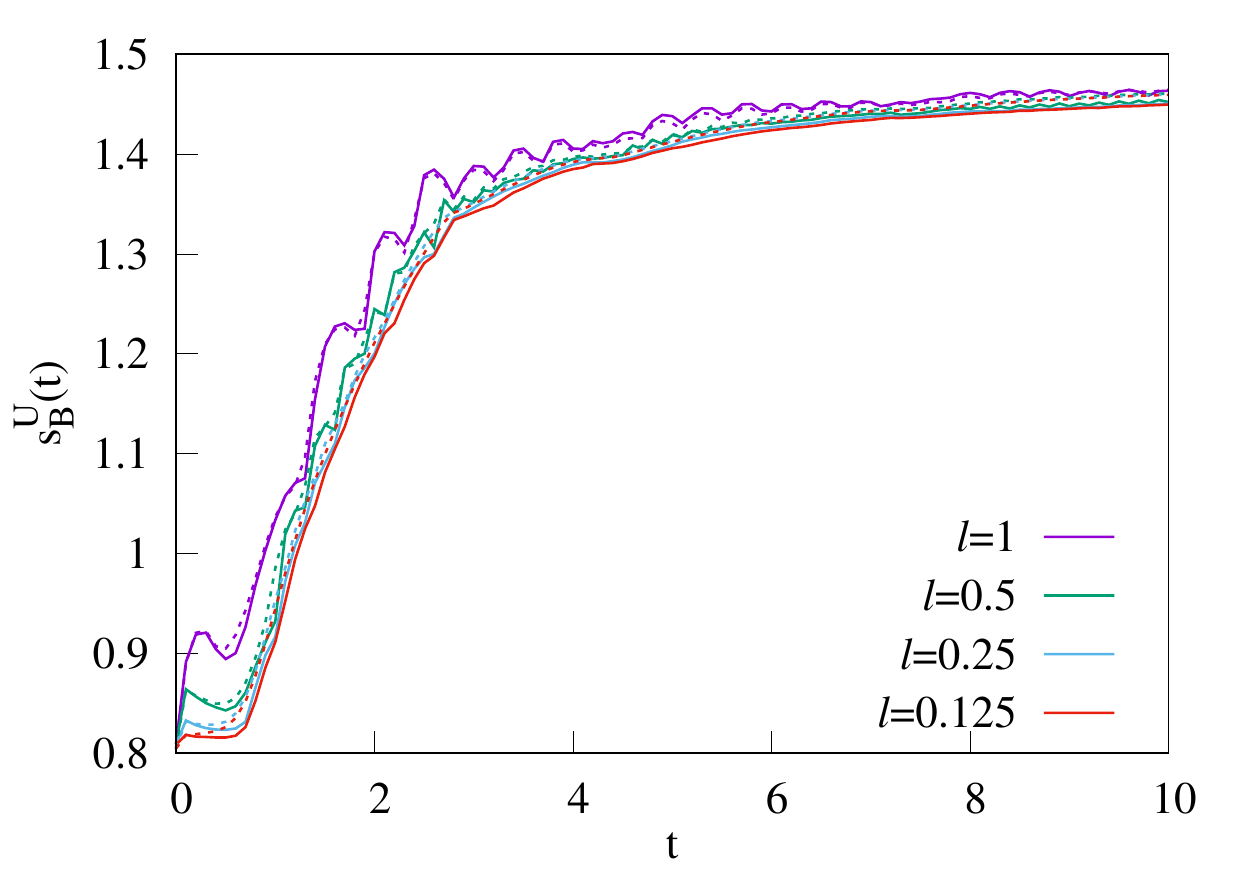}
\put (-180,130) {$\textbf{{\footnotesize(a)}}$}
\includegraphics[width=7.5cm,angle=0]{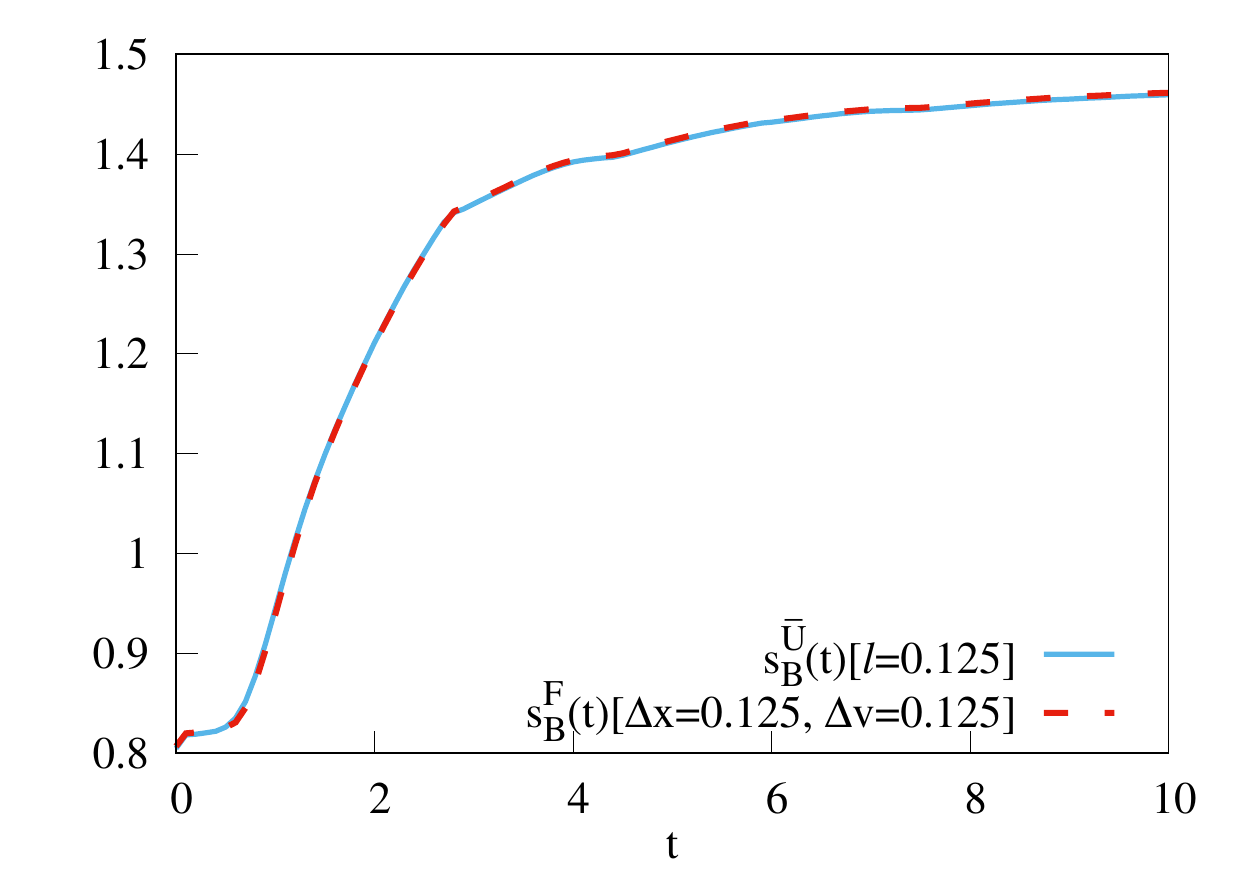}
\put (-180,130) {$\textbf{{\footnotesize(b)}}$}
\caption{(a) We plot entropy per particle $s^U_B(t)$ with time for different cell sizes $\ell$ . The initial condition and parameters are the same as in Fig.~\ref{fig-f_alpha}. Like in Fig.~\ref{fig-sF}, here also we observe a monotonous increment of entropy until it reaches its equilibrium value which is $\ln(2)$ higher than its initial value. The growth rate converges with decreasing cell size $\ell$. The solid and dotted lines correspond to $s^U_B(t)$ and $s^{\bar{U}}_B(t)$ respectively.   (b) Comparison of $s^{\bar{U}}_B(t)$ and $s^F_B(t)$: $s^{\bar{U}}_B(t)$ is calculated for $\ell=0.125$ whereas $s^F_B(t)$ is obtained for $\Delta x=0.125$ and $\Delta v=0.125$. The agreement between two entropies is excellent. The number of particles used for these two plots is $N=4000$.}
\label{fig-sU}
\end{center}
\end{figure}

\begin{figure}
\begin{center}
\leavevmode
\includegraphics[width=14cm]{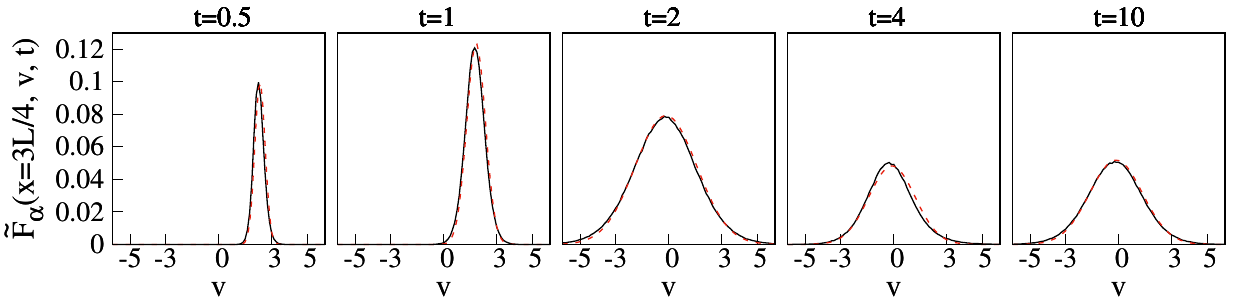}
\caption{Numerical verification of local thermal equilibrium expression in Eq.~\eqref{eq:lte} at different times. The black lines are the marginal velocity distribution $\tilde{F}_\alpha$ and the red dotted lines are the same calculated from Eq.~\eqref{eq:lte} using the data of $\rho$, $u$ and $T$ fields from the cell centered at $x=3L/4$ ($\Delta x=\ell=0.125)$ in  Fig.~\eqref{fig-typicality}. The black lines are obtained using $N=4000$ and averaged over $10^4$ realisations with $\Delta v=0.125$. }
\label{fig-lE-compare}
\end{center}
\end{figure}

%\begin{figure}
%\begin{center}
%\leavevmode
%\includegraphics[width=7.0cm,angle=0]{Entropy-H-function_Scipost.pdf}
%\caption{Comparison of $s^U_B(t)$ and $s^F_B(t)$. $s^U_B(t)$ is calculated for $\ell=0.125$ whereas $s^F_B(t)$ is obtained for $\Delta x=0.125$ and $\Delta v=0.125$. The agreement between two entropies is excellent.}
%\label{fig-compare}
%\end{center}
%\end{figure}

Using the three fields, we obtain the entropy per particle $s^U_B(t)$ from  Eqs.~\eqref{Boltzmann_entropy3} and \eqref{Boltzmann_entropy4} and plot their time evolutions in Fig.~(\ref{fig-sU}a)  for different cell sizes $\ell$. 
 As for the case of $s^f_B(t)$ we find that for the larger cell size $\ell=1$, there is an oscillatory growth with time. This oscillation is related to the motion of the shock front (see Fig.~\ref{fig-xv}). As the shock front enters a new cell of size $\ell$ the overall density fields in the full box tends to become more homogeneous, hence the total entropy starts increasing. This increase will continue till the front reaches the end of the cell which occurs on a timescale $\ell/v_{\rm front}$. However, at this point particles are exiting the cell at a faster rate than they are entering from the preceding cell. Hence, while the inhomogeneity with the next cell decreases, that with the preceding cell increases and we see a decrease of entropy, again till the time the front enters a fresh cell.    
 This argument leads to an oscillation period proportional to $\ell$.  This is consistent with the oscillation period in the $\ell=0.5$ (green) curve being half of that of the curve corresponding to $\ell =1$ (magenta ) as can be seen from Fig.~(\ref{fig-sU}a). These arguments also explain the oscillations observed for $s^f_B$ in Fig.~\eqref{fig-sF}. For smaller values of $\ell$ the growth is 
 monotonic and at long times the net change in entropy is close to the expected value of 
 $\ln(2)$. Further  we observe that the entropy growth rates, for both $s^U_B$ and $s^{\bar{U}}_B$,  at any time converges as we decrease $\ell$.

It is interesting to compare the entropies $s^F_B$ and $s^{\bar{U}}_B$ defined for the two choices of macrostates. It was argued in Sec.~\eqref{sec:intro}  that as a result of local equilibration within small cells, the two entropies should match. In Fig.~(\ref{fig-sU}b) we  plot $s^F_B(t)$ for grid size $\Delta x=0.125$, $\Delta v=0.125$ together with $s^{\bar{U}}_B(t)$ for cell size $\ell=0.125$.  We find an excellent agreement between the two, thus confirming our  expectation from local equilibrium. Note that this is  in sharp contrast to the  equal mass case~\cite{subhadip2021}  where  $s^F_B(t)$ was  oscillatory and its growth rate decreased with decrease of grid size. The presence of local thermal equilibrium is verified in Fig.~\ref{fig-lE-compare}, where we compare the   numerically obtained marginal distribution $\tilde{F}_\alpha$ with the expression in Eq.~\eqref{eq:lte} at a given position $x_\alpha$ for different times.

\subsection{Results for  non-thermal initial condition and mass ratio close to unity}
\label{NT-IC-mr-1}
For the equal mass case $m_1=m_2$, it was shown that there are atypical initial conditions for which the system never equilibrates and the entropy keeps oscillating~\cite{Chakraborti_EM2021}. In this section we explore such initial conditions for our interacting gas, for which we expect monotonic growth and final saturation of the entropy. 
Another interesting question is how the entropy growth curve gets affected when the system is close to the integrable limit {\it i.e.} $m_2/m_1$ close to one. In this section we explore these questions.
%An interesting question is what happens when the masses of alternate particles is close to one. This and the example of atypical initial conditions are briefly discussed in the next section. 

\begin{figure}
\begin{center}
\leavevmode
\includegraphics[width=14cm,angle=0]{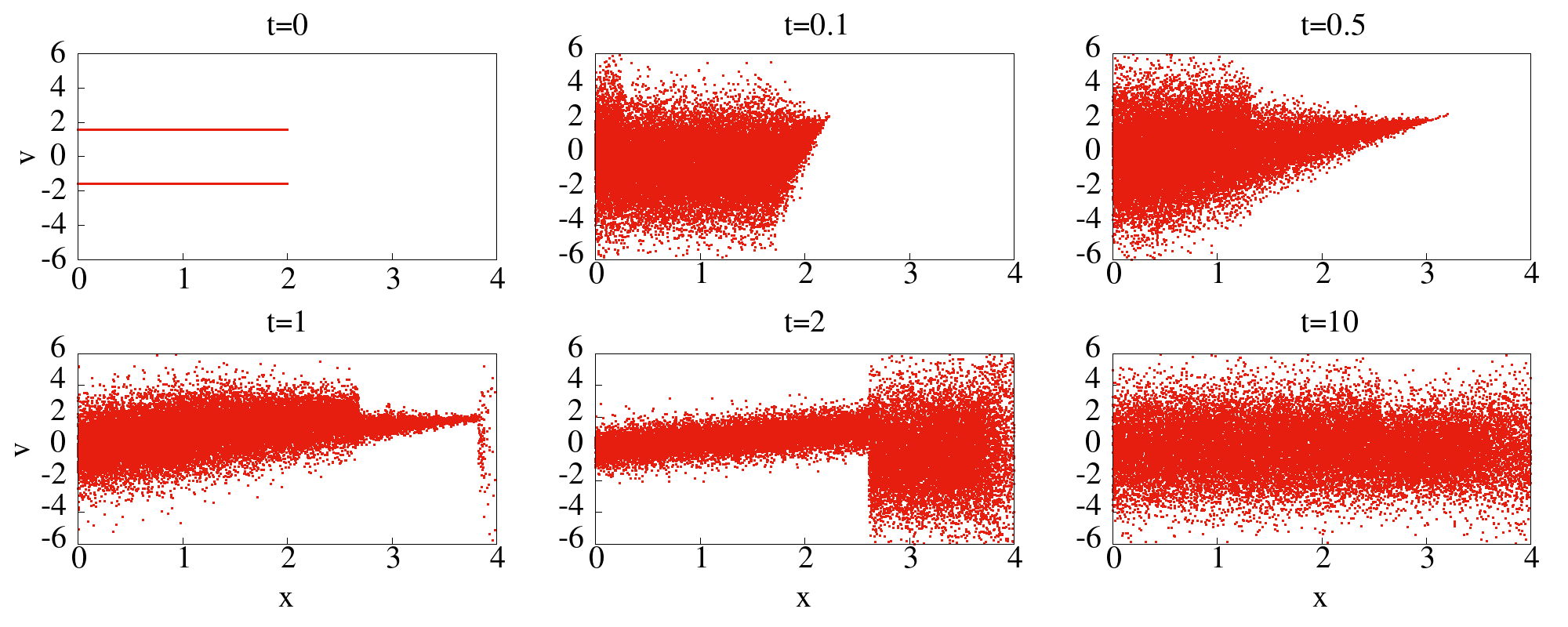}
\caption{Distribution of $N=25000$ particles in the $(x,v)$ plane with binary choice initial velocities. The particles are initially distributed uniformly over the left half $(0,L/2)$ of the system of size $L=4$, with velocities of odd particles  set to $\sqrt{T_0}$ and that of even particles  set to $-\sqrt{T_0}$, where $T_0=2.5$. Due to the presence of interaction, this atypical initial condition  quickly develops to the form of a typical microstate and appears to approach the equilibrium state.}
\label{fig-xv-AM}
\end{center}
\end{figure}

\begin{figure}
\begin{center}
\leavevmode
\includegraphics[width=7.5cm,angle=0]{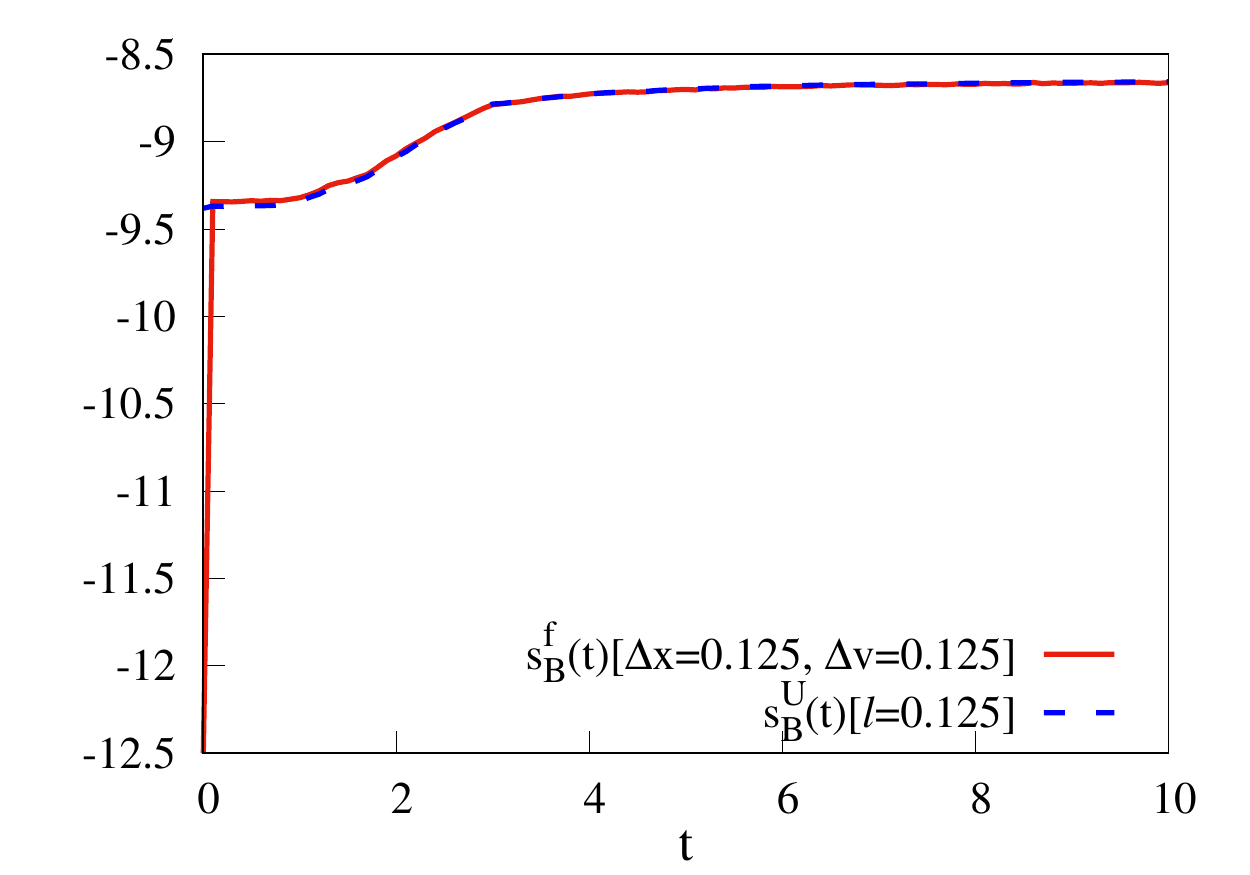}
\put (-175,130) {$\textbf{{\footnotesize(a)}}$}
\includegraphics[width=7.5cm,angle=0]{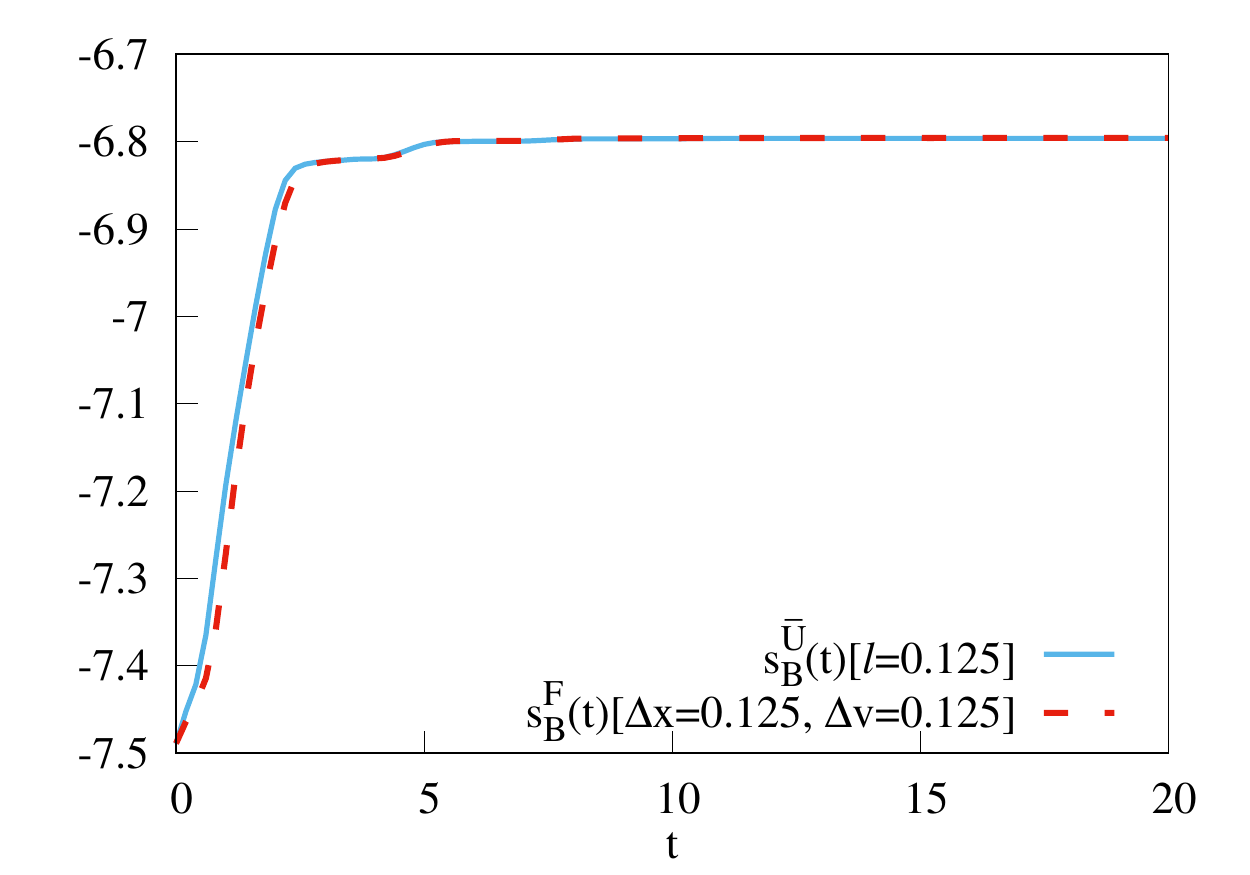}
\put (-180,130) {$\textbf{{\footnotesize(b)}}$}
\caption{(a) Evolution of entropies $s_B^f(t)$ and $s_B^U(t)$ for the binary initial velocity setting as described in Fig.~\eqref{fig-xv-AM}. Since at $t=0$ the microstate is an atypical one and the temperature field is not well defined, there is a significant discrepancy in $s_B^f(t)$ and $s_B^U(t)$. However, this difference goes away quickly and both the entropies grow monotonously towards  equilibrium. 
(b) Plots of the two entropies $s^{\bar{U}}_B(t)$ and $s^F_B(t)$ for the alternate mass gas 
%$m_1=0.975609756$ and $m_2=1.024390244$ 
with  mass ratio $m_2/m_1=1.05$. $s^{\bar{U}}_B(t)$ is calculated for $\ell=0.125$ whereas $s^F_B(t)$ is obtained for $\Delta x=0.125$ and $\Delta v=0.125$. We do not observe any oscillation during the growth, however, the growth rate differs at lower times. This is perhaps due to longer time scales to attain local equilibrium for the case of mass ratio close to unity.}
\label{fig-S-AM}
\end{center}
\end{figure}

{\bf Atypical initial condition}: We consider the case where particles are still  initially uniformly distributed spatially in the left half of the box but    assigned with alternating velocities $+\sqrt{T_0}$ and $-\sqrt{T_0}$. In Fig.~\eqref{fig-xv-AM} we show the evolution of points in $(x,v)$ space for this initial condition. We see that as a result of interactions, the distribution quickly develops into the same form as in Fig.~\eqref{fig-xv} and appears to approach equilibrium.  In the panel (a) of Fig.~\eqref{fig-S-AM} we show the evolution of the corresponding  entropies $s^f_B$ and $s^U_B$ where we find that both quickly converge to the same evolution and show a monotonic growth. The atypicality of the initial state is displayed by the low initial value of the entropy $s^f_B$. Note that for the equal mass case, this initial condition would never reach equilibrium~\cite{Chakraborti_EM2021}.

{\bf Mass ratio close to unity}: To explore the dependence of mass ratio of the alternate mass we consider the case with  mass ratio $m_2/m_1=1.05$. Following the same procedure described in Secs.~\ref{sec:choiceI},\ref{sec:choiceII}, and for the same parameter values, we calculate $s^F_B(t)$ and $s^{\bar{U}}_B(t)$ and plot them in the panel (b) of Fig.~\ref{fig-S-AM}. We observe that even for this mass ratio close to unity, up to a long time, there is no oscillation in both the entropies. Both of them increase monotonically and reach equilibrium value which is $\ln(2)$ higher than its initial value. However, at smaller times the agreement between two entropies is not good possibly because the system takes longer to reach local equilibrium for  $m_2/m_1$ is close to one.

%\begin{figure}
%\begin{center}
%\leavevmode
%\includegraphics[width=8.0cm,angle=0]{Entropy-H-function-mass-ratio-1p05_Scipost.pdf}
%\caption{Plots of two entropies $s^U_B(t)$ and $s^F_B(t)$ for alternate mass $m_1=0.975609756$ and $m_2=1.024390244$ with the mass ratio $m_2/m_1=1.05$. $s^U_B(t)$ is calculated for $\ell=0.125$ whereas $s^F_B(t)$ is obtained for $\Delta x=0.125$ and $\Delta v=0.125$. We does not observe any oscillation during growth, however, the growth rate differs at lower times.}
%\label{fig-mass-ratio1}
%\end{center}
%\end{figure}

\subsection{Connection to thermodynamic entropy and dissipation}
\label{connect-Th}
In the limit of $N \to \infty $ and $\ell \to 0$, the $U$-macrovariables  lead to the following definitions of the  hydrodynamic fields:
\begin{subequations}
\begin{align}
    \rho(x,t)&= \sum_i \la m_i \delta (x -x_i)\ra,\\ p(x,t)&=\sum_i  \la m_i v_i \delta (x -x_i)\ra,\\e(x,t)&=\sum_i  \la m_i \f{v_i^2}{2} \delta (x -x_i)\ra.
\end{align}
\label{eq:exactUeqs}
\end{subequations}
 It can be shown~\cite{Dhar_PRL2001} that they satisfy the exact evolution equations
\begin{subequations}
\label{EulerAll}
\begin{align}
\label{Euler_R_1}
&\partial_t \rho+\partial_x p(x,t)=0, \\
\label{Euler_P_1}
&\partial_t p(x,t) + \partial_x(2e)=0, \\
\label{Euler_E_1}
&\partial_t e  + \partial_x J=0,
\end{align}
\end{subequations}
where we have introduced the local current density field
\begin{equation}
    J(x,t)= \sum_i \la m_i\f{v_i^3}{2} \delta(x-x_i) \ra.
    \label{eq:Jlocal}
\end{equation}
Since $J(x,t)$ is an independent field, the above equations are not really the autonomous equations of hydrodynamics. 

\begin{figure}
\begin{center}
\leavevmode
\includegraphics[width=7.5cm,angle=0]{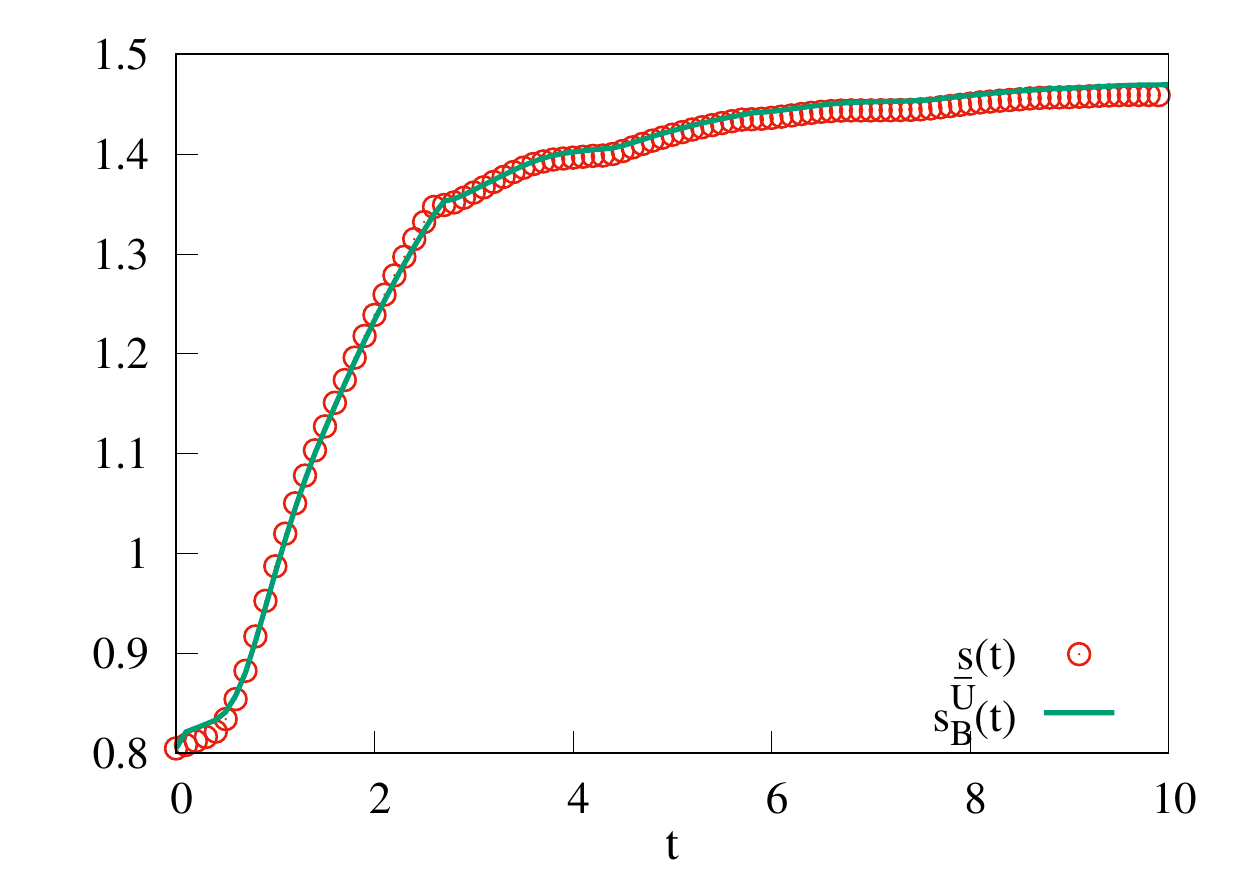}
	\put (-104,145) {$\textbf{{\small(a)}}$}
\includegraphics[width=7.5cm,angle=0]{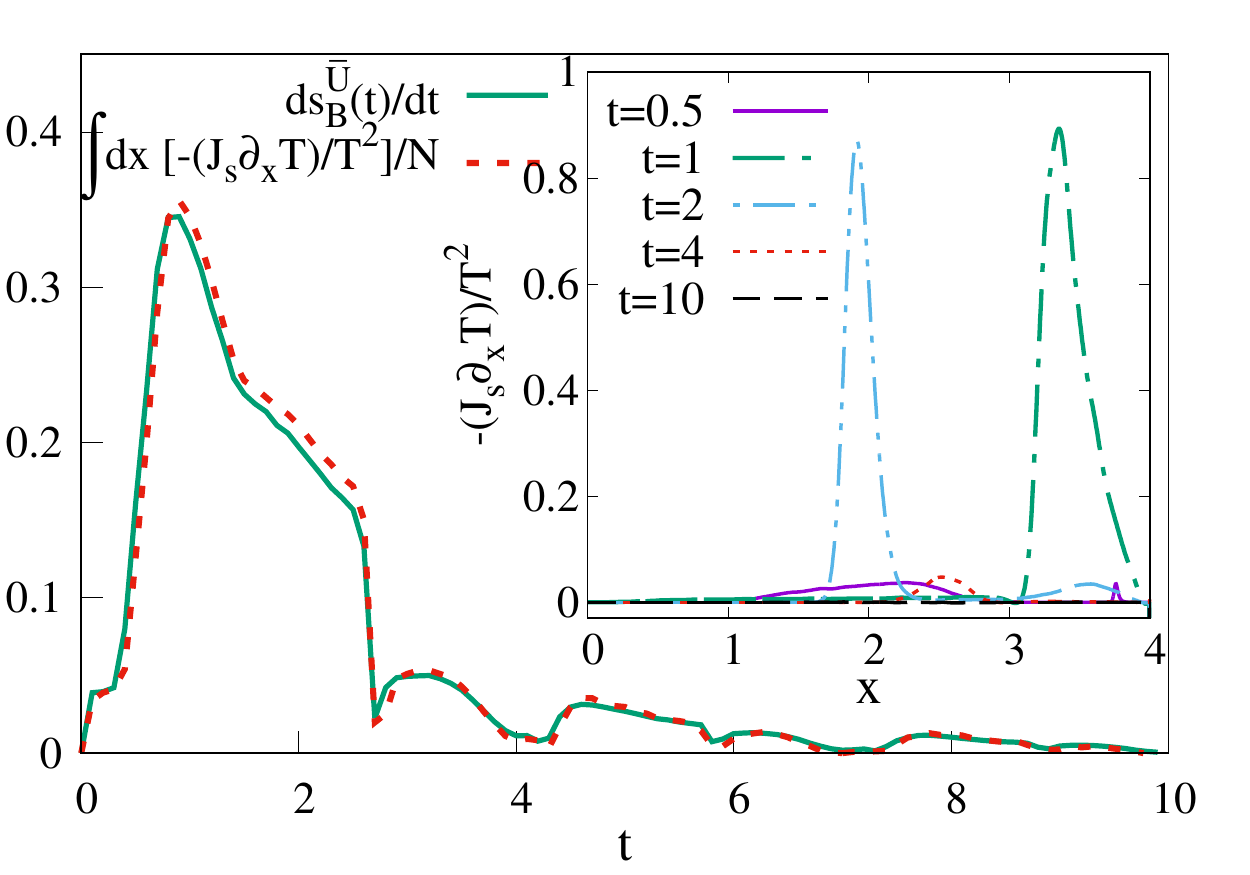}
	\put (-110,145) {$\textbf{{\small(b)}}$}
\caption{(a): Comparison of thermodynamic entropy $s(t)=S(t)/N$ defined in Eq.~\eqref{entropy_production} (red circles) with the Boltzmann entropy $s^{\bar{U}}_B(t)$ (solid green line). Initial values of the entropies were chosen to be consistent with Fig.~\ref{fig-sU}. (b): Comparison of the rate of growth of the two entropies. The inset shows the spatial distribution of entropy production (integrand in Eq.~\eqref{entropy_production}) at different times. We observe that the entropy production at each location is positive and peaked at the locations of the shocks observed in Fig.~\ref{fig-xv}. For all plots the coarse-graining grid size $\ell=0.125$ and the other parameters and the initial conditions are the same as in Fig.~\ref{fig-sU}.}
\label{entropy_compare}
\end{center}
\end{figure}

However, let us write the above equations in a form resembling standard hydrodynamics. For this we define
the  velocity field $u(x,t)=p(x,t)/\rho(x,t)$ and  temperature  field $T(x,t)/2=\tilde{e}(x,t)=\bar{m} e/\rho-\bar{m} u^2/2$ and note the ideal gas relation $P=\rho T/\bar{m}$. We can then write Eqs.~\eqref{EulerAll} in the form:

\begin{subequations}
\label{EulerAll2}
\begin{align}
\label{Euler_R_12}
&\partial_t \rho+\partial_x(\rho u)=0, \\
\label{Euler_P_12}
&\partial_t (\rho u) + \partial_x(\rho u^2+P)=0, \\
\label{Euler_E_12}
&\partial_t e + \partial_x[u (e + P)]=-\partial_x J_s
\end{align}
\end{subequations}
where we have define a new current after subtracting the reversible ``Euler" part as 
\be
J_s(x,t)= J- u  (e + P) \label{eq_W}.
\ee
From  Clausius' definition of entropy $Tds = d\tilde{e} + Pd(1/\rho)$ applied to an infinitesimal volume element, one can get 
the thermodynamic entropy $s(x,t)=s(\tilde{e},\rho)$  per particle \cite{Chakraborti_EM2021}. As the density fields evolve the entropy $s(x,t)$  evolves according to 
\be 
\label{zero}
\frac{Ds}{Dt}=\frac{1}{T} \left[ \frac{D\tilde{e}}{Dt}-\frac{P}{\rho^2}\frac{D\rho}{Dt} \right] = -\f{1}{T}\partial_x J_s,
\ee
%%%%%%%%%%%%%%%%%%%%%%%%%%%%%%%%%%%%%%%%%%
where  $D/Dt=\partial_t + v \partial_x$ is the advective derivative. 
The rate of change of the total entropy in the box is then given by
\bea
\label{entropy_production}
\frac{d S(t)}{dt} = -\int_0^L dx ~ \frac{\partial_x J_s}{T}=-\int_0^L dx \frac{J_s \partial_x T}{T^2}.
\eea
We now compare this with the rate of growth of the Boltzmann entropy $S^U_B$ for the same initial conditions of free expansion studied in Sec.~\eqref{sec:result}. To this end, we numerically compute the current $J_s$ using the  data for the three fields and the local current density $J(x,t)$ in Eq.~\eqref{eq:Jlocal} and use the above formula to compute the growth rate $dS/dt$.  We use a grid size $\ell=0.125$.  For the Boltzmann entropy we use  Eq.~\eqref{Boltzmann_entropy4} and again numerically compute  the entropy growth  rate. In Fig.~(\ref{entropy_compare}a) we compare the time-integrated values of the two entropies   $s(t)=S(t)/N$ and $s_B^{\bar{U}}(t)$,  while in (\ref{entropy_compare}b)  we  compare the  corresponding entropy growth rates. 
In both cases we find very good agreement between the two.  The period of oscillations seen in $ds/dt$ correspond roughly to the time taken by the shock front (see Fig.~\ref{fig-xv}) to travel across the box.
To find the spatial distribution of entropy production at different times, we also plot the integrand in Eq.~\eqref{entropy_production}, $-{J_s \partial_x T}/{T^2}$, in the inset of  Fig.~(\ref{entropy_compare}b)  and find that it is  non-negative and again roughly peaked at the position of the shock front (see Fig.~\ref{fig-xv}).

For a system where Fourier's law is valid one can express $J_s$ in terms of the temperature field as: $J_s = -\kappa \partial_x T$ and then positivity of local entropy  
 growth is guaranteed. For our system, while we do not expect the validity of Fourier's law (due to anomalous heat transport in one dimensions), we have verified that the current $J_s$ still has the properties of a dissipative current {\it i.e.} locally positive entropy production. 

\section{Conclusion}
\label{conclusion}

In this paper, we studied the evolution of Boltzmann's entropy in the alternate mass hard point gas which is a simple model of a one-dimensional  interacting 
gas that however has ideal gas thermodynamics. In the set-up of free expansion inside a box, the Boltzmann's entropy associated with two different macrostates, $f$ and $U$, were studied for single realizations as well as ensembles of initial conditions. We summarize our main findings.
\begin{itemize} 
\item We demonstrated typicality by showing the equivalence of results of the macrostate evolution from single microscopic realizations and from ensembles, in the limit of large $N$. 
\item Unlike the case of equal masses, studied in Ref.~\cite{Chakraborti_EM2021}, here we find that $s^f_B$ converges to a limiting growth curve when one takes the limit $N \to \infty$, $\Delta \to 0$. We discussed the issue of finite size effects and subtleties, in obtaining the true limiting growth curve, by taking this limits in the proper way.
\item As explained in the introduction, the presence of interactions  is expected to lead to local thermal equilibrium, described by Eq.~\eqref{eq:lte},  which in turn would imply that the  limiting growth curves for $s^f_B$ and $s^U_B$ are identical. We provided  a direct numerical verification of Eq.~\eqref{eq:lte} and also a clear demonstration of the equality of $s^f_B(t)$ and $s^U_B(t)$. 
\item We noted interesting features in the evolution of the single particle phase space distribution as well as the hydrodynamic fields. In particular we noted the presence of a shock front which leads to discontinuities in the hydrodynamic fields and oscillatory structures in the entropy growth curve.  
\item Finally we demonstrated the expected equivalence between the growth of Boltzmann's entropy with the thermodynamic Clausius entropy. By writing the thermodynamic entropy growth rate as a spatial integral, we are able to identify a local entropy production rate. We numerically demonstrate it's positivity and show that it is peaked at the shock positions.
\end{itemize}
In conclusion in this work we studied the  role of interactions and non-integrability on the evolution of Boltzmann's entropy during free expansion. Other interesting questions include the evolution of Boltzmann's entropy in interacting integrable systems, such as hard rods and the Toda model,  and in quantum systems.

\section{Acknowledgement}
The authors would like to thank S. Goldstein, D. Huse, J. L. Lebowitz  and   C. Maes for helpful discussions. A. K. also acknowledges the supports of the core research grant no. CRG/2021/002455 and MATRICS grant MTR/2021/000350 from the Science and Engineering Research Board (SERB), Department of Science and Technology, Government of India.  A.K. and A. D. acknowledge support from the Department of Atomic Energy, Government of India, under project no. 19P1112R\&D.  We acknowledge the program -  `Thermalization, Many body localization and Hydrodynamics' (Code: ICTS/hydrodynamics2019/11) at ICTS for enabling discussions related to the project. We also acknowledge the support of the Erwin Schrodinger Institute where various important discussions related to the work were held during the program `Large Deviations, Extremes and Anomalous Transport in Non-equilibrium Systems'.

\bibliographystyle{unsrt}

\bibliography{HPGentropy}

\end{document}